\documentclass{aa}

\usepackage{natbib}
\bibpunct{(}{)}{;}{a}{}{,} 
\usepackage{graphics}   
\usepackage{graphicx}   
\usepackage{epstopdf}
\usepackage{longtable}   
\usepackage{url}      
\usepackage{bm}        
\usepackage[varg]{txfonts}
\usepackage{pdflscape}
\usepackage{supertabular}
\usepackage{hyperref}   
\usepackage[version=3]{mhchem} 

\hypersetup{
    bookmarks=true,         
    unicode=false,          
    colorlinks=true,       
    linkcolor=blue,          
    citecolor=blue,        
    filecolor=blue,      
    urlcolor=blue           
}

\begin{document} 

\title{Constraints on CEMP-no progenitors from nuclear astrophysics}

\author{Arthur Choplin
          \inst{1},
           Andr\'{e} Maeder
           \inst{1},
           Georges Meynet
           \inst{1},
          \and
          Cristina Chiappini\inst{2}
                    }

\titlerunning{Constraints on CEMP-no progenitors from nuclear astrophysics}
\authorrunning{A. Choplin et al.}

\institute{Geneva Observatory, University of Geneva, Maillettes 51, CH-1290 Sauverny, Switzerland
		\and Leibniz-Institut f\"{u}r Astrophysik Potsdam, An der Sternwarte 16, 14482, Potsdam, Germany
		}
\date{Received / Accepted}

\abstract
   {The CEMP-no stars are long-lived small mass stars presenting a very low iron content and overabundances of carbon with no sign or only very weak signs for the presence of s- or r-elements. Although the origin of that abundance pattern is still a matter of debate, it was very likely inherited from a previous massive star, that we shall call here the source star.}
   {We rely on a recent classification of CEMP-no stars arguing that some of them are made of a material processed by hydrogen burning that was enriched in products of helium burning during the nuclear life of the source star. We examine the possibility of forming CEMP-no stars with this material.}
   {We study the nucleosynthesis of the CNO cycle and the Ne-Na Mg-Al chains in a hydrogen burning single zone while injecting the helium burning products $^{12}$C, $^{16}$O, $^{22}$Ne and $^{26}$Mg. We investigate the impact of changing the density and temperature, as well as the injection rate. The nuclear reaction rates involving the creation and destruction of $^{27}$Al are also examined.}
   {$^{14}$N, $^{23}$Na, $^{24}$Mg and $^{27}$Al are formed when injecting $^{12}$C, $^{16}$O, $^{22}$Ne and $^{26}$Mg in the hydrogen burning zone. The $^{12}$C/$^{13}$C ratio is constant under various conditions in the hydrogen burning zone. The predicted [Al/Fe] ratio varies up to $\sim 2$ dex depending on the prescription used for the reaction rates involving $^{27}$Al.
  }
   {The experiments we carried out support the view that some CEMP-no stars are made of a material processed by hydrogen burning, coming from a massive star experiencing mild-to-strong rotational mixing. During its burning, this material was likely enriched in helium burning products. No material coming from  the carbon-oxygen rich core of the source star should be added to form the daughter star, otherwise the $^{12}$C/$^{13}$C ratio would be largely above the observed range of values. 
}


\maketitle

\titlerunning{Spinstars and CEMP-no star}
\authorrunning{Choplin et al.}

\section{Introduction}
\label{intro}

The content of iron at the surface of a star is often used as an indication of the chemical enrichment of its environment. A very small amount of iron relatively to the sun indicates a region similar to the early universe, where only few nucleosynthetic events happened. A way to obtain new clues on stars in the early universe is then to examine the most iron deficient objects. Carbon-Enhanced Metal-poor Stars (CEMP) are a subclass of iron deficient stars, with an excess of carbon relatively to the sun, as well as oxygen and nitrogen in general. Although it can vary a bit from an author to another, the two common criteria defining a CEMP are\footnotemark 
\footnotetext{[X/Y] $=\log_{10}$($N_X / N_Y$) - $\log_{10}$($N_{X\odot} / N_{Y\odot}$) with $N_{X,Y}$ the number density of elements X and Y, $\odot$ denoting the abundances in the sun.}
[Fe/H] $< -1.0$ and [C/Fe] $> 0.7$ \citep{aoki07}.
The range of [C/Fe] covers $\sim 3.5$ dex, from [C/Fe] $=0.7$ to [C/Fe] $=4.26$ for HE1327-2326 \citep{norris13,allen12}. The frequency of CEMP seems to rise toward lower [Fe/H] \citep{lee13}, but also with increasing distance from the Galactic plane \citep{frebel06} or moving from inner to outer halo \citep{carollo12}. The so-called CEMP-no subclass is characterized by its low content in s- or r-elements, contrary to the other sublasses of CEMP: CEMP-s, CEMP-r and CEMP-r/s \citep{beers05}. CEMP-no stars are of particular interest since they dominate at [Fe/H] $\lesssim -3$ \citep{aoki10,norris13}. 

Among the scenarios explored to explain CEMP-no stars, let us mention the "spinstar" scenario \citep{meynet06,meynet10,hirschi07,chiappini13,maeder14} and the "mixing and fallback" scenario \citep{umeda02,umeda05,tominaga14}. The later explains CEMP-no with a model of faint supernovae from Pop III stars with mixing and fallback. The mixing considered in these models occurs just before or during the explosion. The zone of mixing as well as the mass cut\footnotemark
\footnotetext{The mass cut delimits the part of the star which is expelled from the part which is kept into the remnant.}
are free parameters of the models adjusted differently in each star to reproduce observed abundance patterns of CEMP-no stars.

According to the "spinstar" scenario, CEMP-no formed in a region previously enriched by material coming from low metallicity, rotating massive stars. During their nuclear lifetime, spinstars experienced mass loss and strong mixing triggered by rotation. As developed in \cite{maeder14}, although different, these two models appear more complementary than contradictory. Processes like strong internal mixing in the source star, winds or faint supernova may have all happened.
Lately, \cite{takahashi14} presented results based on rotating models with strong fallback (but no mixing in the sense of the works by Umeda \& Nomoto) and tried to deduce the initial rotation of the source stars from comparisons with observed abundance patterns from 3 CEMP-no stars.

\begin{figure*}
   \centering
      \includegraphics[scale=0.52]{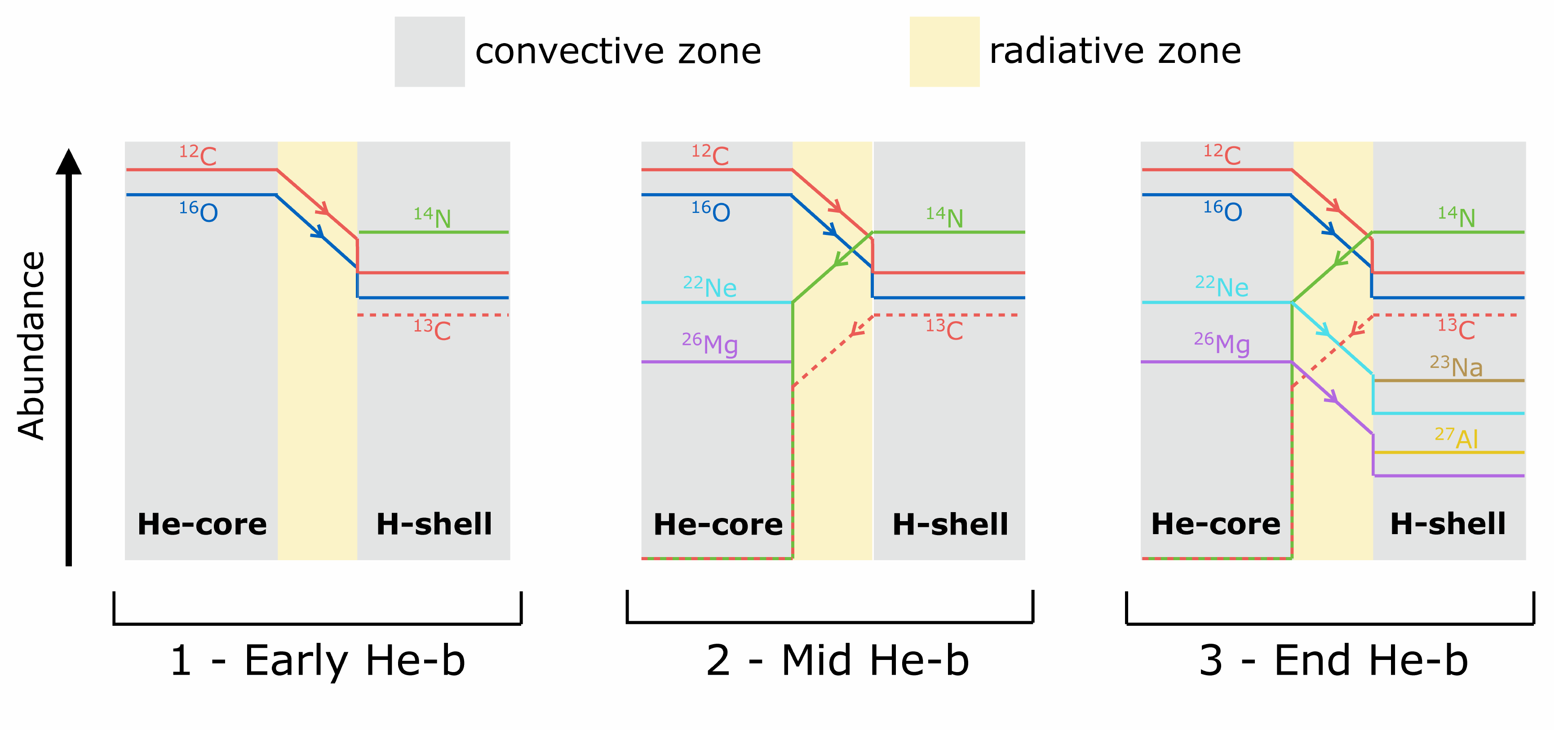}
      \caption{ \label{schema}Schematic view of the 'back and forth' process at work in the spinstar. It occurs during the core helium burning phase and it is an exchange of chemical species between the helium burning core and the hydrogen burning shell.}
\end{figure*}

Recently, \cite{maeder14} proposed the idea that the variety of observed ratios is likely due to material having been processed back and forth by hydrogen and helium burning regions in the spinstar. In other words, these two burning regions are exchanging material between them. These exchanges are triggered by the rotational mixing.
In a first step, the helium burning products diffuse into the hydrogen burning shell. More specifically, $^{12}$C and $^{16}$O synthesized in the helium core diffuse in the hydrogen burning shell, boosting the CNO-cycle and creating primary $^{14}$N and $^{13}$C (see the left panel of Fig.\ref{schema}). In their turn, the products of the hydrogen burning shell (among them $^{14}$N) diffuse back in the helium core. The isotope $^{22}$Ne is synthesized through the nuclear chain $^{14}$N($\alpha,\gamma$)$^{18}$F(,$e^+ \nu_e$)$^{18}$O($\alpha,\gamma$)$^{22}$Ne. 
The isotope $^{26}$Mg can also be synthesized thanks to the reaction $^{22}$Ne($\alpha,\gamma$)$^{26}$Mg (see the middle panel of Fig.\ref{schema}). Also some $^{25}$Mg can be created through the reaction $^{22}$Ne($\alpha,n$)$^{25}$Mg. Neon and magnesium can enter again in the hydrogen burning shell, boosting the Ne-Na and Mg-Al chains and therefore creating sodium and aluminium (see the right panel of Fig.\ref{schema}). Through such back and forth exchanges between the hydrogen and helium burning regions, an all series of isotopes can be formed. The abundances can vary a lot depending on the strength and on the number of those exchanges and thus such models can easily account for the variety of the abundance ratios observed at the surface of CEMP-no stars.
For a given initial mass and rotation rate, the rotational mixing responsible for the exchanges described above is stronger at low metallicity. This effect is mainly due to the higher compactness of low metallicity stars \citep{maeder01}.

\begin{table*}
\scriptsize{
\caption{Type (MS if $T_{eff} \geq 5500$ K and $\log g \geq 3.25$, RGB otherwise), class and abundance data for the CEMP-no stars considered in this work.\label{table:1}} 
\vspace{-0.7cm}
\begin{center}
\resizebox{18.5cm}{!} {
\begin{tabular}{crrrrrrrrrrrrr} 
\hline 
\hline
Star & Type & Class & [Fe/H] & [C/Fe] & [N/Fe] & [O/Fe] & [Na/Fe] & [Mg/Fe] & [Al/Fe] & [Si/Fe] &  [$^{12}$C/$^{13}$C] & A(Li) & Ref.\\ 
\hline 
CS 22945-017 &	MS	& 4++&	-2.52	&	2.28	&	2.24	&      <2.36	&	-	&		0.61	&	-	& 		-	&-1.17 &	<1.51	& 4, 6, 9\\
CS 22949-037 &	RGB	& 4+	&	-3.97&	1.06	&	2.16	&	1.98  	&	2.10 &		1.38	&  0.02	& 		0.77	&-1.35	 &	<0.13	& 7, 9\\
CS 22958-042 &	MS	& 4	&	-2.85	&	3.15	&	2.15	&	1.35	        &	2.85	&		0.32	&	-0.85 & 		0.15	&-1.00	 & 	<1.33& 4, 6, 9\\
CS 29498-043 &	RGB	& 4	&	-3.49	&	1.90	&	2.30	&	2.43  	&	1.47	&		1.52	&	0.34  & 		0.82	&-1.17	 &	<-0.05	& 7, 9\\
CS 30322-023 &	RGB	& 4++&	-3.39	&	0.80	&	2.91	&	0.63  	&	1.04	&		0.80	&	-        & 		0.58	&-1.35 &	<-0.3	 &	1, 4, 5, 6\\
HE 0057-5959 &	RGB	& 2+Na&    -4.08&	0.86	&	2.15	&	<2.77	&	1.98	&		0.51	&      - 	 &		-	&>-1.65	 &	-	&	 7\\
HE 1300+0157 &	MS	& 2	&	-3.75&	1.31	&	<0.71&	1.76  	&	-0.02	&		0.33	&	-0.64 &		0.87	&>-1.47	&	1.06	& 2, 6, 7\\
HE 1310-0536 &	RGB	& 2+	&	-4.15&	2.36	&	3.20	&	<2.80        &	0.19	&		0.42	&	-0.39 & 		<0.25&-1.47 	&	<0.8	& 10, 11	\\
HE 1327-2326 &	MS	& 4++&	-5.76&	4.26	&	4.56	&	3.70  	&	2.48	&		1.55	&	1.23  &		-	&>-1.25	&	<0.62	& 3, 6, 7\\
HE 1410+0213 &	RGB	& 2+	&	-2.52&	2.33	&	2.94	&	2.56	&	-	        &		0.33	&	- 	 & 		-	&-1.47	&	-	& 4, 6 \\
HE 1419-1324 &	RGB	& 4++&	-3.05&	1.76	&	1.47	&	<1.19        &      -	&		0.53	&	- 	 & 		-	&-0.87 & 	-	&	4, 6\\
HE 2331-7155 &	RGB	& 4	&	-3.68&	1.34	&	2.57	&	<1.70        &	0.46	&		1.20	&	-0.38 & 		-	&-1.25 & 	-	&	11\\
SMSS 0313-6708 &	MS	& 4++& <-7.1&	>4.50	&	-	&	-        	&	-  	&		>3.3	&	-	&		-	&  -	& 	0.7	&8\\

\hline
\end{tabular}
}
\end{center}

\vspace{-0.2cm}
\textbf{References}. 1 - \cite{masseron06}; 2 - \cite{frebel07}; 3 - \cite{frebel08}; 4 - \cite{masseron10}; 5 - \cite{masseron12}; 6 - \cite{allen12}; 7 - \cite{norris13}; 8 - \cite{keller14}; 9 - \cite{roederer14a}; 10 - \cite{hansen14}; 11 - \cite{hansen15}.

}

\end{table*}

Putting aside the complexity of stellar models, we realize in this work simple nuclear experiments in order to illustrate the idea of \cite{maeder14} and to constrain the conditions needed in the source stars that would lead to the appropriate nucleosynthesis required to form CEMP-no stars. We study the impact of injecting $^{12}$C, $^{16}$O, $^{22}$Ne and $^{26}$Mg in a hydrogen burning single zone at typical temperatures and densities of the hydrogen burning shell of a $20-60$ $\rm M_{\odot}$ source star model at very low metallicity ($Z = 10^{-5}$). Different sets of nuclear rates are tested for the three main reactions involving $^{27}$Al. We compare our results with a subsample of 5 CEMP-no stars which have a similar metallicity than the metallicity considered in our models and which are, according to \cite{maeder15}, made of a material processed by hydrogen burning, coming from the source star. Note that the active hydrogen burning shell in the source star can be enriched in products of helium burning, as explained previously.
Although limited, these numerical experiments, by focusing mainly on the nucleosynthesis of the problem, 
allow us to explore what just nuclear physics can do and how the results are sensitive to only some nuclear aspects of the problem. As we shall see, even these very simple numerical experiments allow us to obtain very interesting constraints on the sources of CEMP-no stars, constraints that are particularly strong since based on the most simple numerical experiments that we can imagine to do.

In Sect. \ref{sec:2}, we recall briefly the classification of CEMP-no stars made by \cite{maeder15} and select the subsample of CEMP-no stars used in this work. The experiments we carried out are described in Sect. \ref{sec:3} and results obtained in Sect. \ref{sec:4}. Sect. \ref{sec:5} and \ref{sec:6} are dedicated to a discussion about the $^{12}$C/$^{13}$C ratio, the lithium and aluminium abundances. 
In Sect. \ref{sec:7}, we discuss the possible astronomical origin of the CEMP-no stars considered.
Conclusions are given in Sect. \ref{concl}.

\section{The CEMP-no stars of classes 2 and 4}
\label{sec:2}

\cite{maeder15} provided a method to classify the abundance patterns observed at the surface of CEMP-no stars based mainly on two ideas: the first idea is that some material can be exchanged between the hydrogen and helium burning regions inside the star. As already noted in the previous section, the hydrogen burning reactions can transform the material enriched in helium burning products. This will boost the abundance of some isotopes as for instance $^{14}$N. This $^{14}$N can at its turn diffuse into the helium burning region where it is transformed into $^{22}$Ne, and $^{22}$Ne can migrate into the hydrogen burning region, being transformed (at least in part) into $^{23}$Na. Focusing on hydrogen burning regions, we shall speak of secular mixing of \textit{first order} when the CNO cycle will process material enriched by the normal products of helium burning (typically C and O), of \textit{second order} when the CNO cycle will process material enriched in helium burning products resulting from material that was enriched in hydrogen burning products (typically $^{22}$Ne, resulting from $\alpha$-captures on $^{14}$N).
Similar definitions can be made for the helium burning region. 
Just due to these back and forth exchanges, different families of abundance patterns resulting from hydrogen and helium burning and secular mixing of various orders can result.

The second idea is that a second type of mixing can be envisaged: this one occurs between the stellar ejecta, when the nuclear reactions have stopped. We shall call this type of mixing {\it stellar ejecta mixing}. Typically, some CEMP-no stars show signs of being made of material processed by hydrogen and helium burning and then mixed once ejected into the ISM. The material processed by hydrogen burning can result from secular mixing of various orders.

Using such lines of reasoning, \cite{maeder15} divided the CEMP-no class in 5 subclasses.

\begin{itemize}
\item Class 0: the CEMP-no is made of a material processed by hydrogen burning but which was not enriched in helium burning products (no secular mixing, no mixing of the ejecta).
\item Class 1: the CEMP-no is made of a material processed by helium burning but which was not enriched in hydrogen burning products (no secular mixing, no mixing of the ejecta).
\item Class 2: the CEMP-no is made of a material processed by hydrogen burning, which has been enriched in the normal products of helium burning (secular mixing of first order, no mixing of the ejecta).
\item Class 3: the CEMP-no is made of a mixture of ejecta involving material processed by both hydrogen and helium burning. The material processed by hydrogen burning results from a secular mixing of first order, and the material processed by helium burning results from a secular mixing of second order (that means that the material processed by helium burning has been enriched by hydrogen burning products having transformed helium burning products). For instance large amounts of $^{14}$N and $^{13}$C, coming from transformation of $^{12}$C and $^{16}$O, enter by mixing in the helium core. Then, successive $\alpha$-captures on $^{14}$N create some $^{22}$Ne and $^{25,26}$Mg.
\item Class 4: the CEMP-no is made of a material processed by hydrogen burning (no mixing of the ejecta) resulting from secular mixing of second order. This means that the hydrogen burning transforms material that was processed two times by helium burning. Typically, neon and magnesium enter again in the hydrogen shell, boosting the Ne-Na and Mg-Al chains.
\end{itemize}

In each of the classes presented before, refinements are made depending on how advanced is the nuclear burning. For instance, the Mg-Al chain may have more or less acted in the source star so that more or less aluminium has been be created. A sign '+' after the class number indicates a material more processed. A material even more processed is indicated with a sign '++' after the class number. We see that classes 1 and 3 are, at least partly, made of helium burning products while classes 0, 2 and 4 are made of hydrogen burning products. \cite{maeder15} attributed a class to 30 out of 46 CEMP-no stars: 4 belonging to the class 2, 17 to the class 3 and 9 to the class 4.

In the present work, we focus on CEMP-no of classes 0, 2 and 4, i.e. made of a material processed by hydrogen burning that was eventually enriched in helium burning products. A characteristic shared by both stars in classes 2 and 4 (there are not yet observed CEMP-no of class 0) is a relatively low $^{12}$C/$^{13}$C ratio, between 2 and 12 with a mean of 5.1. This value is characteristic of the CNO processing. The other CEMP-no stars have generally a higher $^{12}$C/$^{13}$C (up to 50). 
Part of the helium burning region ($^{12}$C-rich and $^{13}$C-poor) expelled by the source star is used to form the classes 1 and 3, explaining the higher $^{12}$C/$^{13}$C ratios for class 3 CEMP-no stars (there are not yet observed CEMP-no of class 1).

Our subsample is finally made of 13 CEMP-no stars of classes 2 and 4 with a mean [Fe/H] of $-3.9$. Table \ref{table:1} gives the type, the class and the abundances data for the sample of CEMP-no stars considered in this work. We note the class 2+Na for HE 0057-5959. 'Na' stands because of the high [Na/Fe] ratio. The interpretation is the following: owing to a sufficiently high temperature, a significant amount of $^{20}$Ne was synthesized in the helium burning core of the source star. Some of it have diffused in the hydrogen burning shell, boosting the Ne-Na chain and therefore creating some $^{23}$Na.

Note that among those 13 CEMP-no stars, 5 are dwarfs (MS, cf Table \ref{table:1}) and 8 are giants (RGB, cf Table \ref{table:1}), according to the following criteria: the stars with $T_{eff} \geq 5500$ K and $\log g \geq$ 3.25 are dwarfs, the other are giants. The RGB CEMP may have underwent a dredge up event, modifying their surface abundances. Such a dredge up decreases the $^{12}$C surface abundances and increases the $^{13}$C and $^{14}$N surface abundance.
The abundances of O, Ne, Na, Mg and Al elements will not change since the temperature inside the hydrogen burning shell of such a low mass star is likely too low to activate the ON, Ne-Na and Mg-Al cycles.
If we look at the plots [C/Fe] and [N/Fe] versus $\log$ g for the observed CEMP-no stars \citep[c.f.][Fig. 1]{choplin16} we see that the dispersion of the carbon and nitrogen abundances with respect to iron are quite similar for MS and RGB stars. This means that the effect of the first dredge-up does not change significantly the abundance of carbon and nitrogen with respect to the changes related to the dispersion of the initial abundances (about 4 dex). Based on stellar evolution models, \cite{placco14} have determined a correction 
\begin{equation}
\Delta [C/Fe] = [C/Fe]_{\rm ini} - [C/Fe]_{\mathrm{after \, 1DUp}}
\end{equation}
 to apply to the [C/Fe] ratio of 505 Metal-Poor stars in order to recover their initial [C/Fe]. This correction corresponds to the effect of the first dredge-up (if any). Any dredge-up would decrease [C/Fe] so that $\Delta$[C/Fe] $\geq 0$. Note that $\Delta$[C/Fe] $=0$ for MS CEMP since they likely did not experience the first dredge-up. 
5 out of the 8 RGB CEMP considered here belong to this sample; 3 have $\Delta$[C/Fe] $<0.1$, one $\Delta$[C/Fe] $=0.31$ (CS 29498-043) and one $\Delta$[C/Fe] $=0.74$ (CS 22949-037).
These corrections are small compared to the observed range of [C/Fe] ratios.

Also, the fact that the CNO equilibrium value of $^{12}$C/$^{13}$C is obtained at the surface of the MS stars implies that such low values, at least in these stars, cannot be due to a dredge up event. Moreover, the highest $^{12}$C/$^{13}$C belong to HE 1419-1324, a RGB CEMP, showing that the RGB feature is not necessarily associated with a low $^{12}$C/$^{13}$C ratio, as expected by the effect of the dredge up. Correcting the CNO abundances of the evolved CEMP-no stars is of course important in general but 
in the framework of the present work we focus on the range of observed abundances rather than of individual stars, and hence these small corrections have no impact on our conclusions.

\begin{figure}
   \centering
      \includegraphics[scale=0.47]{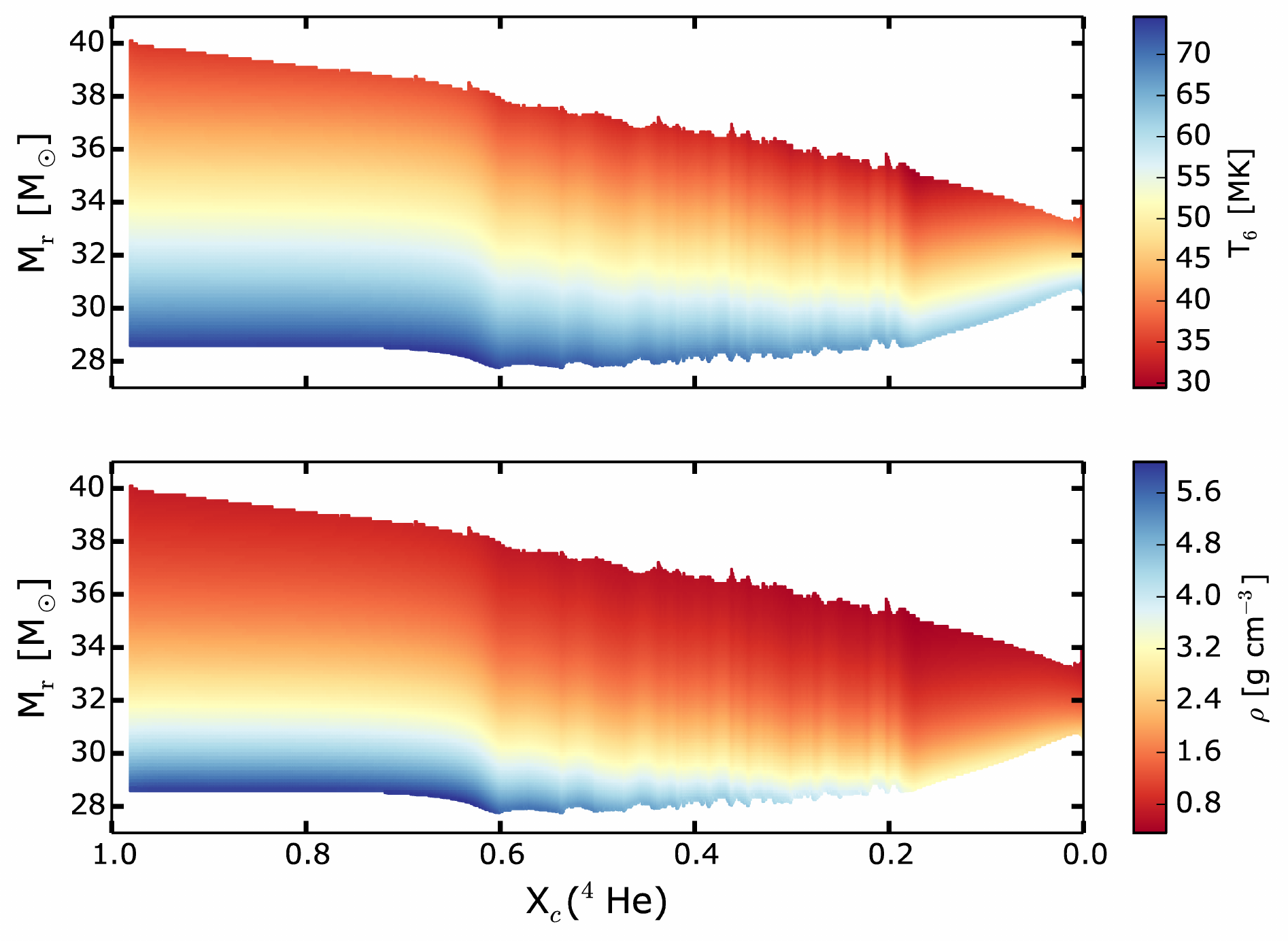}
      \caption{$\rm M_r$ as a function of $X_c(^{4}\rm He)$, the central mass fraction of $^4$He for a rotating 60 $\rm M_{\odot}$ model at a metallicity $Z = 10^{-5}$ (similar to a Kippenhahn diagram). Shown in color is the temperature $T_6$ in MK (upper panel) and the density $\rho$ in g cm$^{-3}$ (lower panel) in the convective hydrogen burning shell, during the core helium burning phase, from $X_c(^{4}\rm He) = 0.98$ to $X_c(^{4}\rm He) = 0$. \label{shell}}
        
\end{figure}

\begin{figure*}
   \centering
      \includegraphics[scale=0.545]{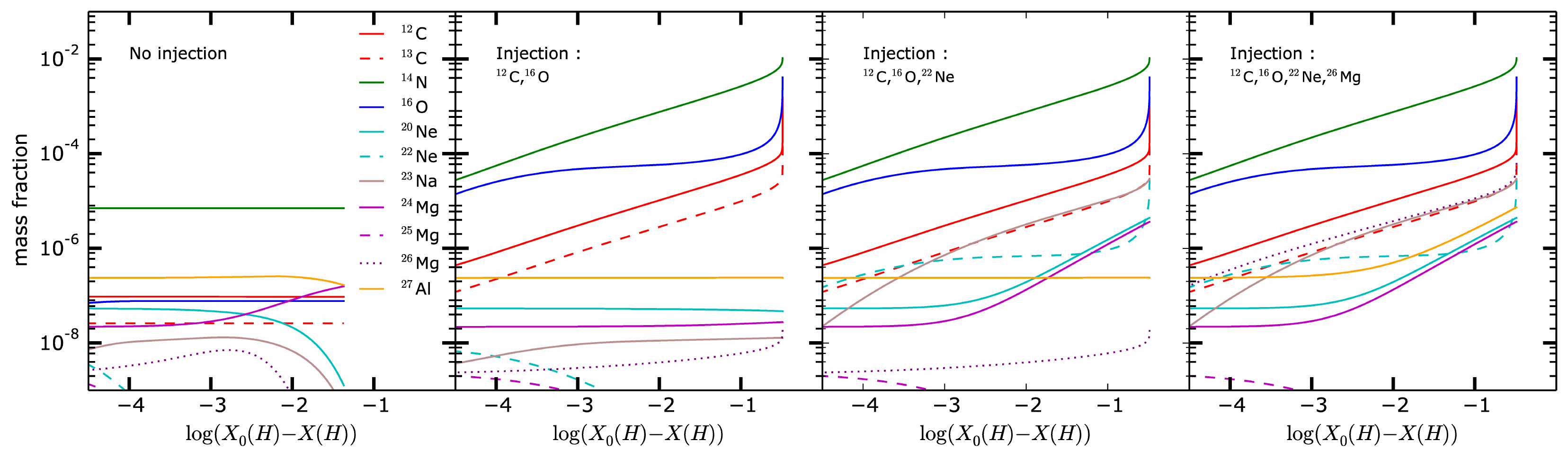}
      \caption{ \label{inj}Abundances in the H-box as a function of the logarithm of $X_0 (H) - X(H)$. The four panels correspond to the four considered cases: when no injection is made, when $^{12}$C and $^{16}$O are injected, when $^{12}$C, $^{16}$O and $^{22}$Ne are injected and when $^{12}$C, $^{16}$O, $^{22}$Ne and $^{26}$Mg are injected. Density and temperature in the box are $\rho = 1$ g cm$^{-3}$ and $T_6= 50$ MK.}
\end{figure*}

\section{Presentation of the experiment}
\label{sec:3}

The conducted experiment consists in injecting products synthesized in the helium burning core of massive stars like $^{12}$C, $^{16}$O, $^{22}$Ne or $^{26}$Mg in a hydrogen burning single zone of 1 $\rm M_{\odot}$ (H-box thereafter) with a constant temperature $T$ and density $\rho$. The H-box reproduces schematically the convective hydrogen burning shell during the core helium burning phase of the source star.

To set the initial conditions in the H-box, we rely on a rotating 60 $\rm M_{\odot}$ model computed with the Geneva code. The initial metallicity is $Z = 10^{-5}$. It corresponds to [Fe/H] $= -3.8$ for the initial mixture we considered ($\alpha$-enhanced). The initial abundances in the H-box are taken from the hydrogen burning shell of this model, at the beginning of the core helium burning phase, when the mass fraction of $^{4}$He in the core $X_c(^{4}\rm He) = 0.98$. Fig. \ref{shell} shows the temperature and density profiles of this model during the core helium burning phase. $\rm M_r$ is the mass coordinate, $X_c(^4 \rm He)$ the central mass fraction of $^4$He. We see from this figure that $T$ and $\rho$ in the convective hydrogen burning shell take values of $30 - 80$ MK and $1 - 5$ g cm$^{-3}$ respectively. For a 20 $\rm M_{\odot}$ model, the ranges of $T$ and $\rho$ are $30 - 60$ MK and $1 - 10$ g cm$^{-3}$. 
As a first step, we fix $T = 50$ MK and $\rho = 1$ g cm$^{-3}$ in the H-box. Different temperatures and densities in the H-box are discussed in a second time. The simulation is stopped either when the hydrogen in the box is exhausted (when the mass fraction of hydrogen in the H-box $X(^{1}\rm H)<10^{-8}$) or when the time $t$ exceeds 10 Myr. Note that depending on the stellar model chosen for setting the initial abundances in the H-box, we can have a slightly different initial chemical composition in the H-box. This will depend on the chemical composition of the model in its hydrogen burning shell, at the core helium burning ignition ($X_c(^{4}\rm He) = 0.98$). The CNO abundances in the hydrogen burning shell at the core helium burning ignition do not change significantly from a 20 $\rm M_{\odot}$ to a 60 $\rm M_{\odot}$ model. The abundances of neon, sodium, magnesium and aluminium do vary a bit more. It is due to the difference of temperature that implies a slightly different nucleosynthesis in the hydrogen burning shell, at the very beginning of the core helium burning phase. 

Regarding the nuclear reaction rates, we took the ones used in the Geneva code \citep[see][]{ekstrom12}. Those rates are mainly from \cite{angulo99} for the CNO-cycle but almost all rates for Ne-Na Mg-Al chains are from \cite{hale02}. Only $^{20}$Ne($p,\gamma$)$^{21}$Na and $^{21}$Ne($p,\gamma$)$^{22}$Na are taken from \cite{angulo99} and \cite{iliadis01}, respectively. Note that the final abundance of $^{26}$Al in the H-box is added to the one of $^{26}$Mg since $^{26}$Al is a radioactive isotope ($t_{1/2} = 7.17 \times 10^{5}$ yrs).

In order to reproduce schematically the diffusion of $^{12}$C, $^{16}$O, $^{22}$Ne and $^{26}$Mg from the helium core to the hydrogen shell, we inject a constant mass per year in the H-box, coming from a reservoir composed only of the considered specie ($^{12}$C, $^{16}$O, $^{22}$Ne or $^{26}$Mg). We consider injection rates of $10^{-10}$, $10^{-8}$ and $10^{-6}$ $\rm M_{\odot}$ yr$^{-1}$ for $^{12}$C and $^{16}$O and $10^{-12}$, $10^{-10}$ and $10^{-8}$ $\rm M_{\odot}$ yr$^{-1}$ for $^{22}$Ne and $^{26}$Mg. More details about the method for injecting the species and the justification of the adopted injection rates are given in the Appendix.
Four cases are tested in the present experiment:
\begin{itemize}
\item no injection is made in the H-box,
\item $^{12}$C and $^{16}$O are injected,
\item $^{12}$C, $^{16}$O, and $^{22}$Ne are injected,
\item $^{12}$C, $^{16}$O, $^{22}$Ne and $^{26}$Mg are injected.
\end{itemize}

Note that in a real star, the mass is conserved and thus any injection into the hydrogen burning shell implies that some matter has to diffuse out from that region.
In complete stellar models, the elements that are more abundant in the hydrogen burning shell than in one of the two adjacent regions will diffuse out in the region(s) where this element is less abundant. However to keep the model as simple as possible we do not consider that complication here.  The present work can be seen as a numerical experiment and not as an attempt to model in all details what happens in stars.
Indeed, the most important gradients of abundances are those coming from the difference in the abundances between the helium core and the hydrogen burning shell. Diffusion from the helium core to the hydrogen burning shell is therefore clearly the dominant feature, blurring all the other diffusion processes. In complete stellar models, injection of nitrogen into the helium core occurs dominantly by convection when the helium core slightly grows in regions left over by the hydrogen burning shell, not by diffusion from the hydrogen to the helium burning region.
Also, as we will see, the results of our simple box model well fit qualitatively the results obtained from complete stellar models following in a consistent way the mixing of the elements.
Thus we are quite confident that our simple approach grasp the essential of the process.

Let us also mention that the origin of the iron is not investigated in the present work. By choosing a non-zero metallicity, we assume that the small iron content observed at the surface of the CEMP-no stars is already present in the source star (and in the H-box). Our sample of classes 2 and 4 CEMP-no stars is expected to be made of a material processed by hydrogen burning that comes from the source star, i.e. of the outer layers of this star. The iron abundance in those outer layers is likely not affected by the nucleosynthesis and stays equal to its initial value. As a consequence, a comparison of the models with the observed ratios like [C/Fe] or [N/Fe] can be made, provided that the iron content in the models (the [Fe/H] ratio) is similar to the iron content of the CEMP-no stars. In our models, [Fe/H] $=-3.8$ so that the observed CEMP-no stars around this value can be consistently compared with the models.

\begin{figure*}
   \centering
      \includegraphics[scale=0.55]{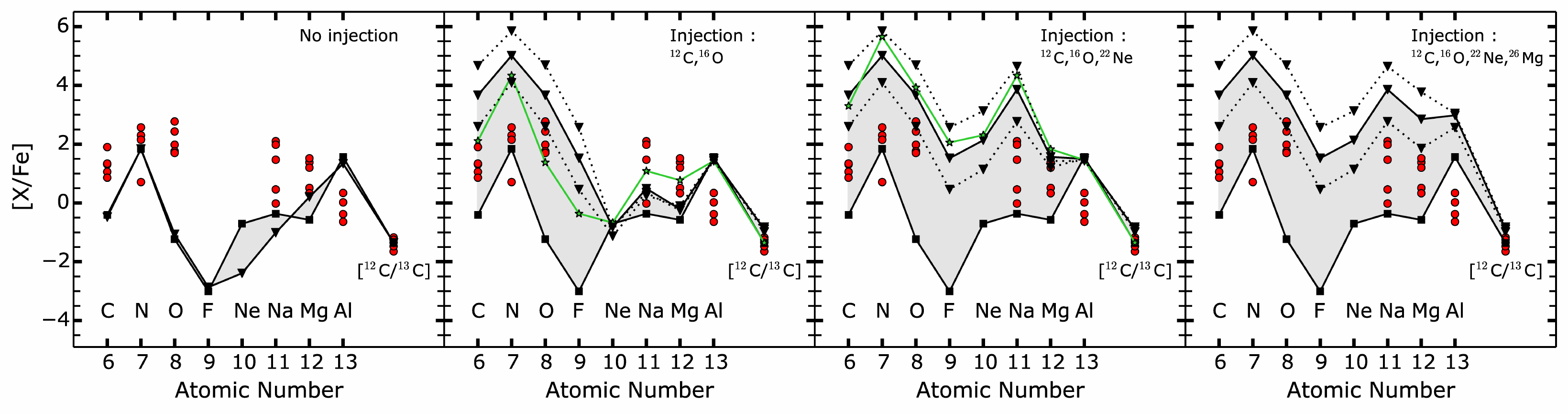}
      \caption{\label{xfe}[X/Fe] ratios in the H-box for the four considered cases. The [$^{12}$C/$^{13}$C] ratio is also shown (i.e. $^{12}$C/$^{13}$C ratio relatively to the sun, in logarithm). The red points show the CEMP-no stars of Table \ref{table:1} with [Fe/H] $=-3.8 \pm 0.3$. The abundance patterns in the box at $t=0$ are represented by black lines with squares and the patterns at the end of the simulation by black lines with triangles. Lower and upper black dotted lines represent the final composition in the H-box when the rates of injection are divided and multiplied by $10^2$, respectively. The green pattern on the second (third) panel shows the composition in the hydrogen burning shell at the end of the core helium burning phase of a complete 20 $\rm M_{\odot}$ stellar model at 30\% (70\%) of the critical velocity on the ZAMS. Density and temperature in the box are $\rho = 1$ g cm$^{-3}$ and $T_6= 50$ MK.
      }   
\end{figure*}

\section{Results of the experiment}
\label{sec:4}

Fig. \ref{inj} shows the mass fraction of elements as a function of $\log(X_0(H) - X(H))$ (i.e. the logarithm of the initial mass fraction of hydrogen minus the current hydrogen mass fraction in the H-box) for the four cases presented in Sect. \ref{sec:3}. The initial mass fraction of $^{1}$H in the box is equal to 0.33, the temperature and density are set to $T=50$ MK and $\rho = 1$ g cm$^{-3}$ and the injection rates are $10^{-8}$ $\rm M_{\odot}$ yr$^{-1}$ for $^{12}$C and $^{16}$O and $10^{-10}$ $\rm M_{\odot}$ yr$^{-1}$ for $^{22}$Ne and $^{26}$Mg.

\begin{itemize}
\item When no injection is made (first left panel of Fig. \ref{inj}), only 4.28 $\times$ 10$^{-2}$ of hydrogen (in mass fraction) is consumed at the end of the limited time, which we fixed at 10 Myr. No transient regime is seen for the CNO elements since the CNO-cycle is already at equilibrium at $t=0$. The mass fraction of sodium, $X(^{23}$Na) first increases slightly due to the effect of the Ne-Na chain. It finally drops, like $X(^{20}$Ne), in favor of $X(^{24}$Mg) owing to the reaction $^{23}$Na($p,\gamma$)$^{24}$Mg. At $T_6 = 50$, since the 
$^{24}$Mg($p,\gamma$)$^{25}$Al reaction is weak, it does not succeed in transforming efficiently the $^{24}$Mg synthesized so that the $^{24}$Mg abundance increases. A little amount of $^{27}$Al is destroyed when little $^{1}$H remains in the shell and is transformed either in $^{28}$Si through $^{27}$Al($p,\gamma$)$^{28}$Si, or in $^{24}$Mg through $^{27}$Al($p,\alpha$)$^{24}$Mg, both channels being almost equal at this temperature.
\item Injecting $^{12}$C and $^{16}$O boosts the CNO cycle and creates primary $^{14}$N and $^{13}$C (second left panel of Fig. \ref{inj}). The CNO elements being more and more abundant, more and more hydrogen is burnt and it is finally exhausted after $\sim$ 0.8 Myr. The reactions in the CN-cycle are fast so that an equilibrium is almost instantaneously reached for $^{12}$C, $^{13}$C and $^{14}$N. 
Since the reaction $^{16}$O($p,\gamma$)$^{17}$F is much slower, the injected $^{16}$O accumulates before being destroyed into $^{17}$F. Those two regimes can be seen on the second left panel of Fig. \ref{inj}: the curve showing $^{16}$O first increases until $\log(X_0(H) - X(H)) \sim - 3$ (accumulation) and then becomes flatter for $-3 < \log(X_0(H) - X(H)) < - 1$ ($^{16}$O destruction becomes important).
No permanent regime is attained for $^{16}$O. $^{12}$C and $^{16}$O rise dramatically at the end since the hydrogen is almost exhausted: the CNO cycle works less and less, implying an accumulation of the injected $^{12}$C and $^{16}$O. Regarding the other elements, we see that $^{22}$Ne decreases in favor of $^{23}$Na but the duration of the simulation is too short in this case for the Ne-Na and Mg-Al chains to operate significantly.
\item Doing the same experiment but injecting also some $^{22}$Ne allows the synthesis of $^{23}$Na and $^{24}$Mg through the reactions $^{22}$Ne($p,\gamma$)$^{23}$Na and $^{23}$Na($p,\gamma$)$^{24}$Mg (see Fig. \ref{inj}, third left panel). Also some $^{20}$Ne is created when the reaction $^{23}$Na($p,\alpha$)$^{20}$Ne occurs. At this temperature and for the selected rates, the ($p,\alpha$) channel is 1.6 higher than the ($p,\gamma$) channel so that $^{23}$Na is almost equally destroyed in $^{20}$Ne and $^{24}$Mg. The reaction $^{24}$Mg($p,\alpha$)$^{25}$Al is too slow to activate the Mg-Al chain. Since neon, sodium and magnesium stay much less abundant than the CNO elements, the hydrogen is not burnt significantly quicker than in the case 2 described just before, when injecting only $^{12}$C and $^{16}$O. 
\item The last case is when injecting $^{26}$Mg as well (right panel of Fig. \ref{inj}). The difference with the previous case is the increase of $^{27}$Al by $\sim$ 2 dex due to the direct transformation of $^{26}$Mg into $^{27}$Al thanks to a proton capture. For the temperature, density and time of simulation considered, the $^{27}$Al destruction through either the ($p,\alpha$) or ($p,\gamma$) channels is not significant. 
\end{itemize}

Fig. \ref{xfe} shows the [X/Fe] ratios for the four cases presented in Fig. \ref{inj}. 
Plotted are the initial abundance pattern in the H-box (lines with squares) and the final one (lines with triangles). The grey area corresponds to the range of values covered during the simulation. The dotted lines correspond to the final patterns in the H-box when the rates of injection are divided and multiplied by 100 respectively.  
The red points represent the observed ratios in CEMP-no stars of similar metallicities ([Fe/H] $=-3.8 \pm 0.3$) as the metallicity considered in our numerical experiment ([Fe/H] $=-3.8$). As mentioned in Sect. \ref{sec:3}, it is important to focus on the CEMP-no that have similar metallicities than the one in the models since the iron abundance in our model is an initial condition that is not modified by the nuclear reactions occurring in the H-box. Any change of the initial iron abundance in the models will produce a shift of the [X/Fe] ratios predicted by the models.

Without injection, the abundance pattern in the H-box does not change very much and all the [X/Fe] ratios (except [N/Fe] and [Al/Fe]) stay below the observed values. It is not surprising since the observed values correspond to classes 2 and 4 while the present experiment (no injection) would rather correspond to the CEMP-no of class 0. This class is made of a material processed only by hydrogen burning and where no mixing occurred between the hydrogen and helium burning regions. 
Injecting some $^{12}$C and $^{16}$O enhances the corresponding [X/Fe] ratios, as well as [N/Fe] (Fig. \ref{xfe}, second left panel). 
Increasing or decreasing the rate of injection by a factor of 100 (dotted lines) changes the final pattern in the box but not dramatically: from the lower to the upper dotted line, the rate of injection is multiplied by 4 dex while the difference in [X/Fe] values does not exceed 2 dex (for C, N and O). This is because when injecting more $^{12}$C and $^{16}$O in the H-box, the hydrogen is burnt more rapidly so that hydrogen exhaustion occurs earlier. We have here a negative feedback process: increasing the rate of injection increases the amount of injected species but at the same time reduces the available time for injecting those new chemical species in the H-box. This explains qualitatively why injecting $^{12}$C and $^{16}$O at a rate $10^4$ higher does not lead to an increase of 4 dex of the final [C/Fe], [N/Fe] and [O/Fe] ratios. 
The third case shows enhancements of [Ne/Fe], [Na/Fe] and [Mg/Fe] ratios: protons captures on the injected $^{22}$Ne create $^{23}$Na and then $^{24}$Mg.
The final patterns of the fourth case present enhancements of [Mg/Fe] and [Al/Fe] ratios compared to the case 3: due to the injection of $^{26}$Mg, the Mg-Al chain is boosted, hence creating some $^{27}$Al.

\begin{figure*}
   \centering
      \includegraphics[scale=0.85]{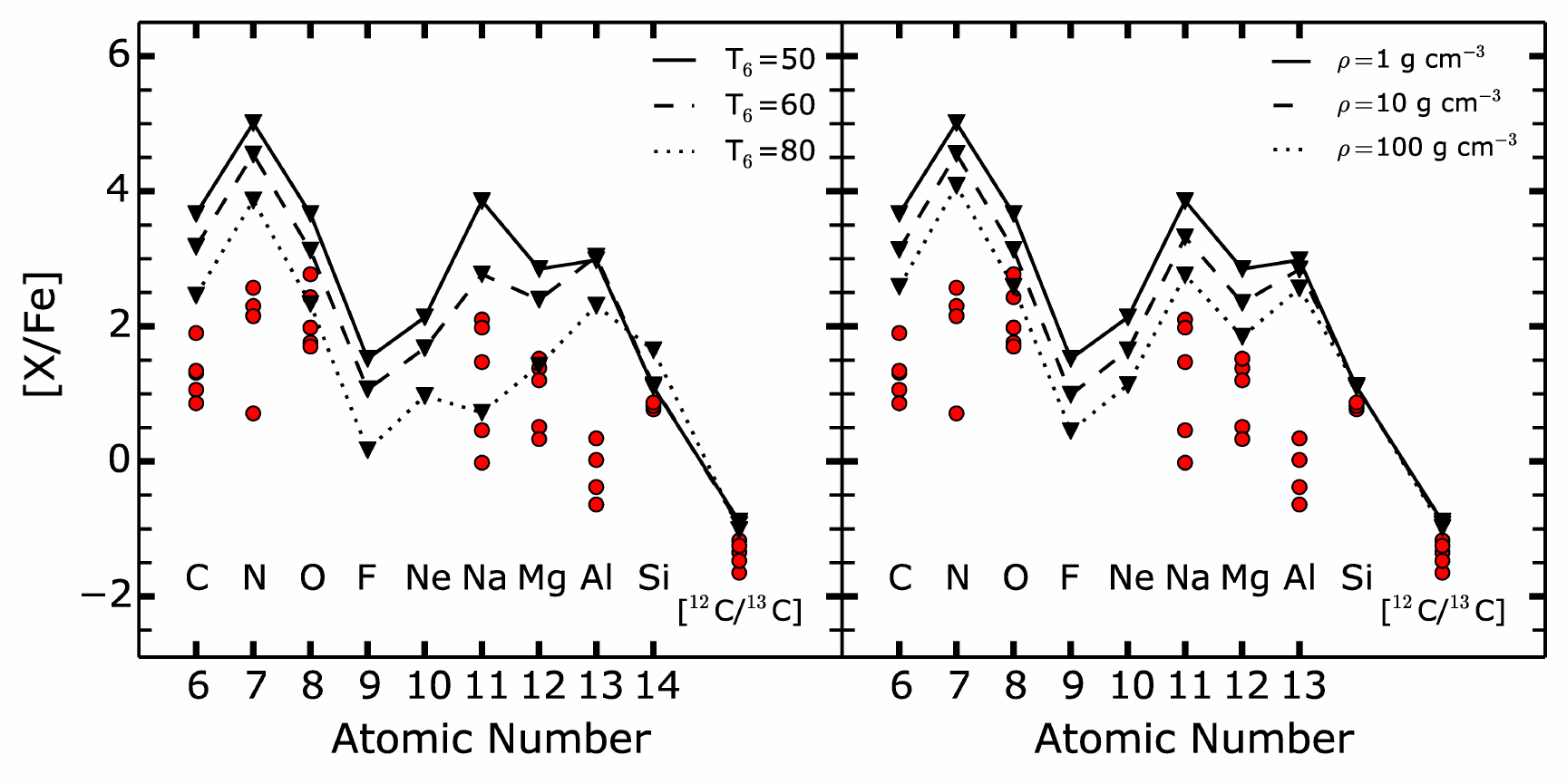}
      \caption{\label{rhoT} \textit{Left}:  same as figure \ref{xfe} but for 3 different temperatures in the H-box. Also the [Si/Fe] ratio is shown. Only the final abundance patterns in the box are plotted. The injected species are $^{12}$C, $^{16}$O, $^{22}$Ne and $^{26}$Mg. Density is unchanged (1 g cm$^{-3}$) and the injection rate is 10$^{-8}$ $\rm M_{\odot}$ yr$^{-1}$ for $^{12}$C and $^{16}$O and 10$^{-10}$ $\rm M_{\odot}$ yr$^{-1}$ for $^{22}$Ne and $^{26}$Mg). \textit{Right}: same as the left panel but for 3 different densities in the H-box. The temperature is set to 50 MK.}
\end{figure*}

Changing the temperature and the density in the H-box leads to the results presented in Fig. \ref{rhoT}. We tested temperatures of $T_6 = $ 50, 60 and 80 MK at a constant density $\rho = $ 1 g cm$^{-3}$ (left panel) and densities of  $\rho =$ 1, 10 and 100 g cm$^{-3}$ at a constant temperature $T_6 = 50$ MK (right panel). For both cases, the injected species are $^{12}$C, $^{16}$O, $^{22}$Ne and $^{26}$Mg and the injection rates are 10$^{-8}$ $\rm M_{\odot}$ yr$^{-1}$ for $^{12}$C and $^{16}$O and 10$^{-10}$ $\rm M_{\odot}$ yr$^{-1}$ for $^{22}$Ne and $^{26}$Mg. In addition to the other [X/Fe] ratios, [Si/Fe] is also shown.
Increasing either the temperature or the density leads to lower [X/Fe] ratios at hydrogen exhaustion (except however for the [Si/Fe] ratio at $T_6 = 80$ MK). When raising the density for instance, the rates of the nuclear reactions increase, allowing a quicker synthesis of the chemical species than at lower densities. At the same time, the hydrogen is burnt more rapidly so that hydrogen exhaustion occurs earlier, letting less time to inject new species. Summing those two opposite effects finally leads to lower [X/Fe] ratios at hydrogen exhaustion.
For the same reasons, similar results are found when varying the temperature, although the dependance of the nuclear rates on temperature is much stronger than the dependance on the density.
This is the reason why the various temperatures spanning a relatively small range of values (see Fig. \ref{rhoT}) change the [X/Fe] ratios much more significantly than the various densities, that cover yet a much larger range of values.

The initial [Si/Fe] ratio in the box is about 1 and it is little affected by changes of temperature and density. However, increasing the temperature up to $T_6 = 80$ MK leads to about 0.5 dex more silicon at the end. The first reason is that the nuclear rates associated to Mg, Al and Si are generally 3-4 dex higher at $T_6=80$ MK than at $T_6=50$ MK, allowing the synthesis of some $^{28}$Si. The second reason is the following: $^{27}$Al is destroyed either to form $^{28}$Si thanks to the $^{27}$Al($p,\gamma$)$^{28}$Si reaction, either to form $^{24}$Mg owing to the reaction $^{27}$Al($p,\alpha$)$^{24}$Mg. 
When $^{27}$Al is destroyed into $^{24}$Mg, the Mg-Al chain operates and synthesizes again $^{27}$Al. On the opposite, $^{27}$Al is definitely destroyed when it is transformed into $^{28}$Si.
At $T_6 = 50$ and 60 MK, both channels are roughly equal. At $T_6 = 80$ MK, the ($p,\gamma$) channel is $\sim 40$ times higher than the ($p,\alpha$) one. This tends to reduce Mg and Al and to increase Si.

Also shown on Fig. \ref{xfe} and \ref{rhoT} is the [$^{12}$C/$^{13}$C] ratio, i.e. $\log$($^{12}$C/$^{13}$C) - $\log$($^{12}$C/$^{13}$C)$_{\odot}$. The isotopic ratio in the Sun is taken from \cite{lodders03}. Whatever the case, the [$^{12}$C/$^{13}$C] ratio does not vary more than 0.5 dex, always staying around -1.5. This is because the equilibrium value is quickly reached when the CNO cycle operates. After each injection of $^{12}$C, the equilibrium ratio is reached again quasi instantaneously compared to the current timestep.
Especially, [$^{12}$C/$^{13}$C] reaches similar final values under all the explored temperatures and densities. When the CNO cycle operates, the [$^{12}$C/$^{13}$C] equilibrium ratio is indeed almost temperature and density independent.

We see that the observed [X/Fe] ratios are best covered when injecting $^{12}$C, $^{16}$O and $^{22}$Ne (see the grey area in the third panel of Fig. \ref{xfe}). We note however that the models always give too high values for the [Al/Fe] ratios (this point will be discussed on Sect. \ref{sec:6}). 
For all the other ratios, our very simple numerical experiment confirms the need for some mixing between the helium and hydrogen burning zones in the source star to explain the general pattern observed in CEMP-no stars of classes 2 and 4. The fact that the injection of $^{22}$Ne seems to be needed supports the view that a strong mixing might be at work in the source star: $^{22}$Ne can enter in the hydrogen burning shell if $^{12}$C and $^{16}$O have first diffused in the hydrogen burning shell, but also if the created $^{14}$N have entered at its turn in the helium burning core.
For most of the ratios, however, we note that the grey region is wider than the ranges covered by the observations.
We think that this is not a very serious problem when considering that CEMP-no stars are not made
of pure hydrogen shell material. For being used to form new stars, this matter needs to be ejected either by winds or at the time of the supernova. In this process, the region where the CNO-cycle was active (hydrogen shell) will be mixed with other layers of the star as well as eventually with some interstellar medium.
For instance, any mixing with the outer layers of the star where the iron has the same abundance as in the hydrogen shell but where the CNO abundances are smaller, close to their initial values ($\sim 10^{-5}$ for a model with $Z=10^{-5}$) will shift the nitrogen abundance downward.
Therefore, in order to obtain the observed nitrogen abundances in CEMP-no stars, it is likely needed that much higher
abundances are reached in the hydrogen burning shell. While our box experiments provide some interesting constraints about the nuclear processes
that might be needed to reproduce the peculiar abundance patterns of CEMP-no stars, only
the computation of complete stellar models with the account for the ejection mechanism (both through winds and through the
supernova explosion) and for some possible mixing with the circumstellar material can provide abundances ratios that
might be compared with the observed ratios in CEMP-no stars. This has to be kept in mind when interpreting the comparison shown in Fig. \ref{xfe}.

Note also that injecting some $^{26}$Mg raises the [Al/Fe] ratio far above the observed range.
In stellar models, $^{26}$Mg comes from the transformation of $^{22}$Ne which occurs at the very end of the core helium burning phase. Thus, $^{26}$Mg could be injected (if any) only at the very end or after the core helium burning phase, leaving little time for nuclear burning to transform this $^{26}$Mg into $^{27}$Al in the hydrogen burning shell. In that respect, the present numerical experiments injecting $^{26}$Mg regularly all along the burning of the hydrogen shell clearly overestimates what occurs in real stars. Thus, the no $^{26}$Mg injection hypothesis is by far not unrealistic.

It is interesting to note that detailed stellar models are qualitatively well enough reproduced by this simple one zone model. The two green patterns on Fig. \ref{xfe} show the [X/Fe] ratio in the hydrogen burning shell of a complete stellar model at the end of the core helium burning phase. The abundances are taken in the hydrogen shell, where the energy released by hydrogen burning is the highest. The two stellar models are 20 $\rm M_{\odot}$, $Z=10^{-5}$ stars computed at 30\% (second panel) and 70\% (third panel) of the critical velocity on the ZAMS. It corresponds to an initial equatorial velocity of 280 and 610 km s$^{-1}$ respectively. Increasing the initial velocity can be modeled, in the single zone model, by increasing the injection rate and injecting some $^{22}$Ne in addition to the $^{12}$C and $^{16}$O. By comparing the two green patterns, it is also interesting to see the strong impact of the initial rotation on the [X/Fe] ratios in the hydrogen burning shell at core helium exhaustion. This shows that the stellar rotation at low metallicity is likely a non-negligible process.

One point that deserves more discussion is the constant value found for the $^{12}$C/$^{13}$C ratio under various conditions. What can be learnt from this ratio? How could it be used in stellar evolution models? The lithium content at the surface of the CEMP-no stars can give interesting constraints as well. Also a deepest investigation on the [Al/Fe] ratio seems worthwhile, because of the discrepancy we found between models and observations. 
Which conditions or assumptions will favor a lower [Al/Fe] ratio, closer to the observed values?

\section{Dilution with material processed by helium burning and with initial ISM}
\label{sec:5}

In the previous section we have investigated the secular mixing between the hydrogen and helium burning regions, during the nuclear life of the source star. We will now discuss the mixing events that can happen outside from the star, when the nuclear reactions are not active anymore.
Different kinds of materials can be ejected by the source star in the ISM. Let us consider two of them: (i) the material that was processed by hydrogen burning and (ii) the one that was processed by helium burning. Outside from the star, those materials can be mixed together and/or with the ISM. We distinguish 2 kinds of dilutions:

\begin{itemize}
\item the dilution between the material processed by hydrogen burning and the one processed by helium burning. This corresponds to a mixing between different parts of the star. By mixing we mean mixing of the ejecta (or \textit{stellar ejecta mixing} as defined in Sect. \ref{sec:2}), hence outside of the star, without nuclear reactions.
\item the dilution of the whole material ejected by the source star (no matter if it was processed by hydrogen or helium burning) with the initial ISM, in which the source star formed. This is a mixing of the stellar ejecta with the initial ISM.
\end{itemize}

Based on considerations on the $^{12}$C/$^{13}$C ratio and the lithium abundance, we investigate the possibility of constraining those two kinds of mixing events.

\subsection{The $^{12}$C/$^{13}$C ratio to constrain the amount of ejecta processed by helium burning}

The $^{12}$C/$^{13}$C ratio of the selected CEMP-no stars gives a strong constraint on the kind of material needed to form those stars. As we see in Fig. \ref{xfe}, this ratio is very close to the one found in a CNO processed material. We see from Fig. \ref{xfe} and \ref{rhoT} that injecting some $^{12}$C in a hydrogen burning region does not change the $^{12}$C/$^{13}$C ratio, under various densities, temperatures and rates of injection. However, if switching off the nuclear reactions (this is what happen when the material is ejected from the source star) and mixing the CNO processed material with part of the region processed by helium burning, this ratio will no longer stay around the observed values.

\begin{figure}
   \centering
      \includegraphics[scale=0.47]{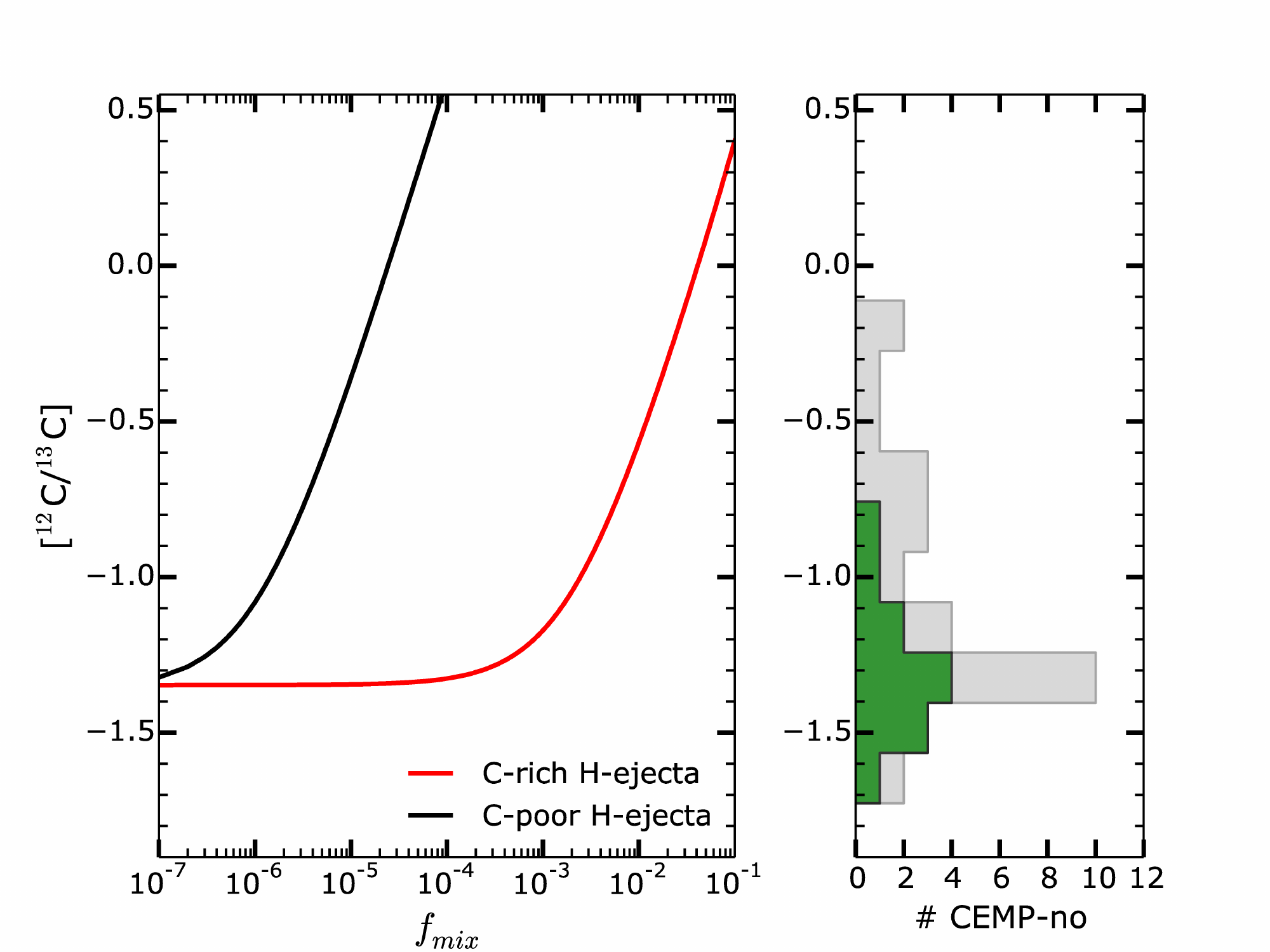}
      \caption{\label{C12C13} \textit{Left}: [$^{12}$C/$^{13}$C] ratio as a function of the mixing factor $f_{mix}$, which represents the fraction of material processed by helium burning mixed with the material processed by hydrogen burning. Two compositions are tested for the material processed by hydrogen burning: a C-rich composition (red line) and a C-poor composition (black line). \textit{Right}: the green histogram shows the distribution of the CEMP-no stars used in this work. They belong to the class 2 or 4. The grey histogram shows all known CEMP-no stars with a measured $^{12}$C/$^{13}$C ratio.}
\end{figure}

This point can be illustrated with a simple experiment. One can mix together the two kind of material ejected by the source star: the ejecta that was processed by hydrogen burning and the one by helium burning (H-ejecta and He-ejecta hereafter). The [$^{12}$C/$^{13}$C] ratio can be followed as adding more and more He-ejecta to the H-ejecta. The left panel of Fig. \ref{C12C13} shows [$^{12}$C/$^{13}$C] as a function of $f_{mix}$ defined as the fraction of He-ejecta added to the H-ejecta. For instance, $f_{mix}=10^{-2}$ means 1\% of He-ejecta with 99\% of H-ejecta. 
We tested two compositions for the H-ejecta. The first one has mass fraction of $^{12}$C and $^{13}$C equal to 1.11 10$^{-7}$ and 3.10 10$^{-8}$. It is called \textit{C-poor H-ejecta}. Those mass fractions are the ones in the H-box at $t=0$. The second mixture has a $^{12}$C mass fraction of 2.01 10$^{-4}$ and a $^{13}$C one of 5.45 10$^{-5}$. We call it \textit{C-rich H-ejecta}. Those values correspond to the mass fractions of $^{12}$C and $^{13}$C in the H-box, a few time before hydrogen exhaustion, when $^{12}$C is injected. 

\begin{itemize}
\item The \textit{C-poor H-ejecta} corresponds to the low rotational mixing case. Few $^{12}$C has diffused from the helium core to the hydrogen shell so that the mass fractions of $^{12}$C and $^{13}$C in the hydrogen burning shell stay around their initial value, i.e. around $10^{-7}$. When this part of the star is expelled, we get an ejecta poor in carbon.
\item The \textit{C-rich H-ejecta} corresponds to the strong rotational mixing case. A lot of $^{12}$C has diffused in the hydrogen burning shell, raising its abundance and the one of $^{13}$C far above their initial values. At the time of the ejection, this material is enriched in carbon compared to the previous case.
\end{itemize}

From what regards the carbon abundances in the He-ejecta, we took characteristics values in the helium burning core. We set X$_{Heb}$($^{12}$C) = 0.1 and X$_{Heb}$($^{13}$C) = 0 as a correct approximation.

We see on the left panel of Fig. \ref{C12C13} that mixing 1\% (i.e. $f_{mix} = 10^{-2}$) of the He-ejecta with 99\% of a C-rich H-ejecta leads to a [$^{12}$C/$^{13}$C] of $-0.6$ which lies above the range of observed values for the CEMP-no stars of classes 2 and 4 (see the green histogram on left panel of Fig. \ref{C12C13} and Table \ref{table:1}). If no rotational mixing has occurred between the helium burning core and hydrogen burning shell of the source star, the H-ejecta is C-poor. In this case, mixing the same amount of He-ejecta as before with the H-ejecta material leads to a [$^{12}$C/$^{13}$C] ratio of 2.6, far above the observed range.
It seems that whatever the C-richness of the H-ejecta, hence the amount of helium products that entered in the hydrogen burning shell during the life of the source star, the final contribution of the material processed by helium burning coming from the the source star should be null to form the CEMP-no stars of classes 2 and 4.
This strongly support the idea that the CEMP-no stars of classes 2 and 4 are only made of the hydrogen envelope of the source star.
More generally, since $^{12}$C/$^{13}$C is highly sensitive to the burning region considered (it is low for a hydrogen burning and high for a helium burning region), this ratio could be used to constrain the mass cut of spinstar models at the time of the supernova: the mass cut could be chosen in order to reproduce the observed $^{12}$C/$^{13}$C ratio of the considered CEMP-no star.

The grey histogram in the right panel of Fig. \ref{C12C13} shows the distribution of all observed [$^{12}$C/$^{13}$C] ratios at the surface of CEMP-no stars. It contains the 13 CEMP-no stars of Table \ref{table:1} plus 15 other CEMP-no stars with a measured $^{12}$C/$^{13}$C ratio. Some CEMP-no stars (not considered in this work) have higher [$^{12}$C/$^{13}$C] ratios, suggesting the need for a small amount of material processed by helium burning to form them. We see however that the amount of material processed by helium burning needed should stay small in any case ($f_{mix} \lesssim 0.05$): mainly the hydrogen envelope of the progenitor should be used to form the CEMP-no stars. Nevertheless, we note that some CEMP-no stars have only a lower limit for the [$^{12}$C/$^{13}$C] ratio (3 out of 12 in our subsample). An accurate determination of this ratio for those stars would be interesting to validate the previous statement regarding those stars.

\subsection{A(Li) to constrain the dilution with the initial ISM}

In the previous section, we discussed the mixing of the stellar ejecta: the regions processed by hydrogen and helium burning ejected by the source star can be mixed together, when the nuclear burning has stopped. The following discussion is related to the mixing (or the dilution) of the whole stellar ejecta with the initial ISM, in which the source star formed.

The lithium is an interesting element to obtain pieces of informations on the amount of ISM that should be mixed with the source star ejecta to form the CEMP-no stars. The abundance of lithium A(Li) in the pristine ISM is equal to 2.72 according to \cite{cyburt08}. It is totally destroyed in massive stars. As a consequence, as soon as the ejecta of the massive source star is mixed with the ISM, the abundance of lithium is raised (in the mixture made of initial ISM and source star ejecta). If one assumes that the lithium content at the surface of the CEMP-no star reflect the one in the cloud where it formed, then the higher the lithium content at the surface of the CEMP-no star, the more the source star ejecta was diluted with the ISM. The dilution factor for mixing the progenitor ejecta with the ISM can be chosen in order to obtain the observed A(Li) value of the considered CEMP-no star.

A difficulty is that the lithium at the surface of the CEMP-no star can be depleted by internal mixing processes in the CEMP-no star itself. However, such processes might not be able to explain the low content of lithium observed at the surface of some CEMP-no stars. \cite{meynet10} pointed out that the maximal depletion predicted by the models of \cite{korn09} (1.2 dex) is unable to account for the A(Li) value observed at the surface of HE 1327-2326 (A(Li) $<$ 0.62, see Table \ref{table:1}). According to the models of \cite{korn09}, one would indeed expect an observed value A(Li) = $2.72 - 1.2 = 1.52$ at minimum, i.e. the WMAP content minus the maximal predicted depletion factor. We see that the depletion mechanism has difficulties to account for the lowest observed A(Li) values. The alternative for HE 1327-2326 is that it formed with a Li-poor material.

The A(Li) values (or upper limits) for 9 of the considered CEMP-no stars are shown on Table \ref{table:1}.
The lower panel of Fig. 3 in \cite{meynet10} shows the dilution factor $\rm M_{ISM} / M_{eje}$ vs. A(Li).
According to this figure and if we consider that the lithium was not depleted by the CEMP-no stars themselves, the dilution factor should be less than $\sim 0.1$ for the stars considered in this work. The highest dilution factor being for CS 22945-017, which has A(Li) $<$ 1.51.
The final mass fraction of $^{12}$C, $^{14}$N and $^{16}$O in the H-box when injecting $^{12}$C and $^{16}$O is at least $10^{-3}$ (see Fig. \ref{inj}). The mass fraction of the CNO elements in a $Z=10^{-5}$ ISM is about $10^{-6}$. 90\% of $10^{-3}$ with 10\% of $10^{-6}$ gives $\sim$ $10^{-3}$. The dilution is not playing a significant role in that case.

Let us suppose now that the lithium was depleted by the CEMP-no stars. We take the maximum depletion factor (1.2 dex) from \cite{korn09} and we add it to the observed A(Li) in order to get the initial A(Li) value, before the depletion process. The two highest A(Li) belong to CS 22945-017 ($<2.71$) and CS 22958-042 ($<2.53$). A(Li) being close to the WMAP value for CS 22945-017, it would imply a high dilution factor. However, for CS 22958-042 and all the other considered CEMP-no stars, the dilution factor should be less than $\sim2$. There is still not enough ISM for the dilution to have a significant effect, except however for CS 22945-017. One must stay cautious on those simple statements about the dilution between ejecta and ISM but in the framework of our simple model, we see that the dilution with the initial ISM might play only a limited role. This is because the metal abundances in the region processed by hydrogen burning are much higher than the ones in the initial ISM and because the dilution factors derived from the lithium abundance are small in most of the cases.

\begin{figure*}
   \centering
   \begin{minipage}[c]{.49\linewidth}
       \includegraphics[scale=0.6]{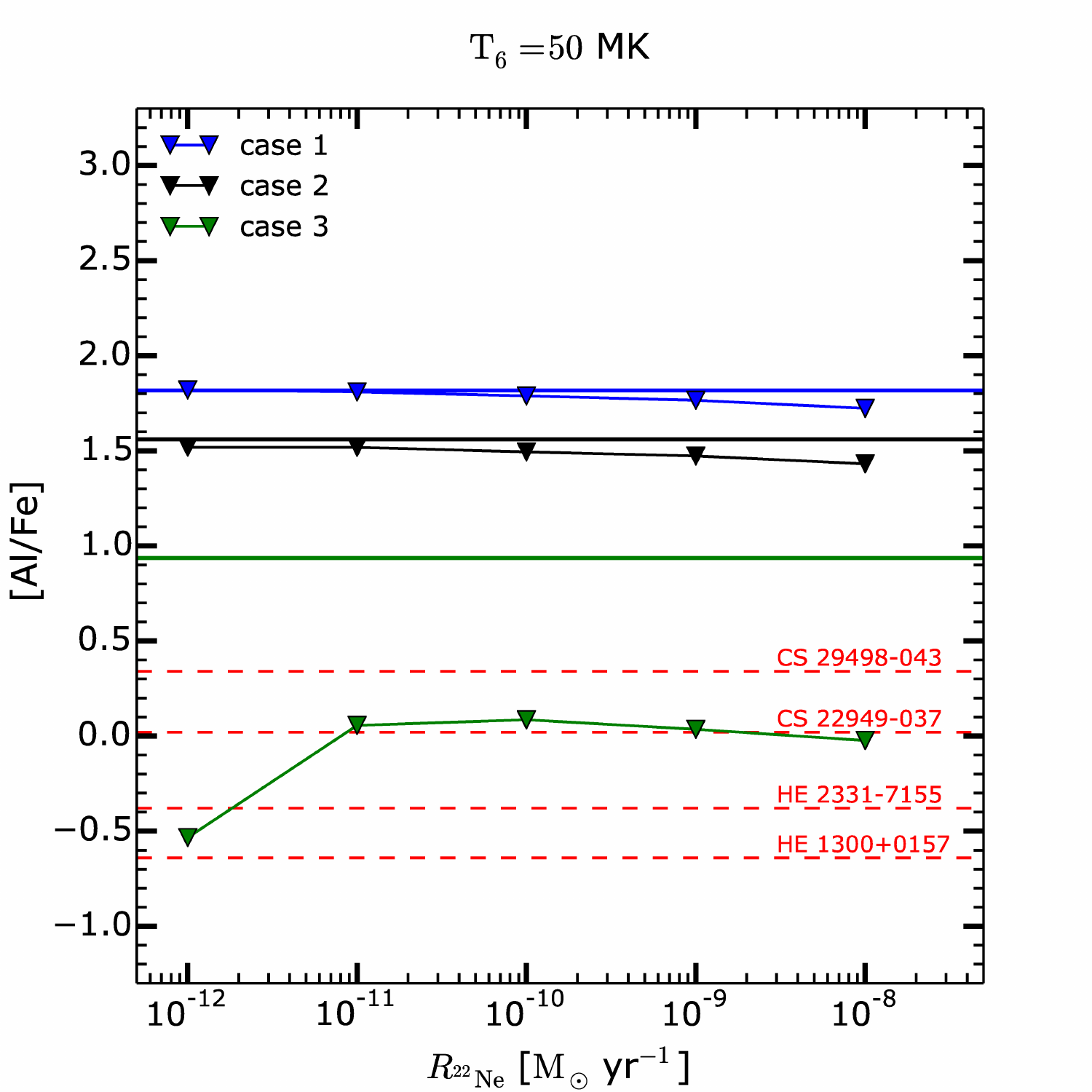}
   \end{minipage} \hfill
   \begin{minipage}[c]{.49\linewidth}
       \includegraphics[scale=0.6]{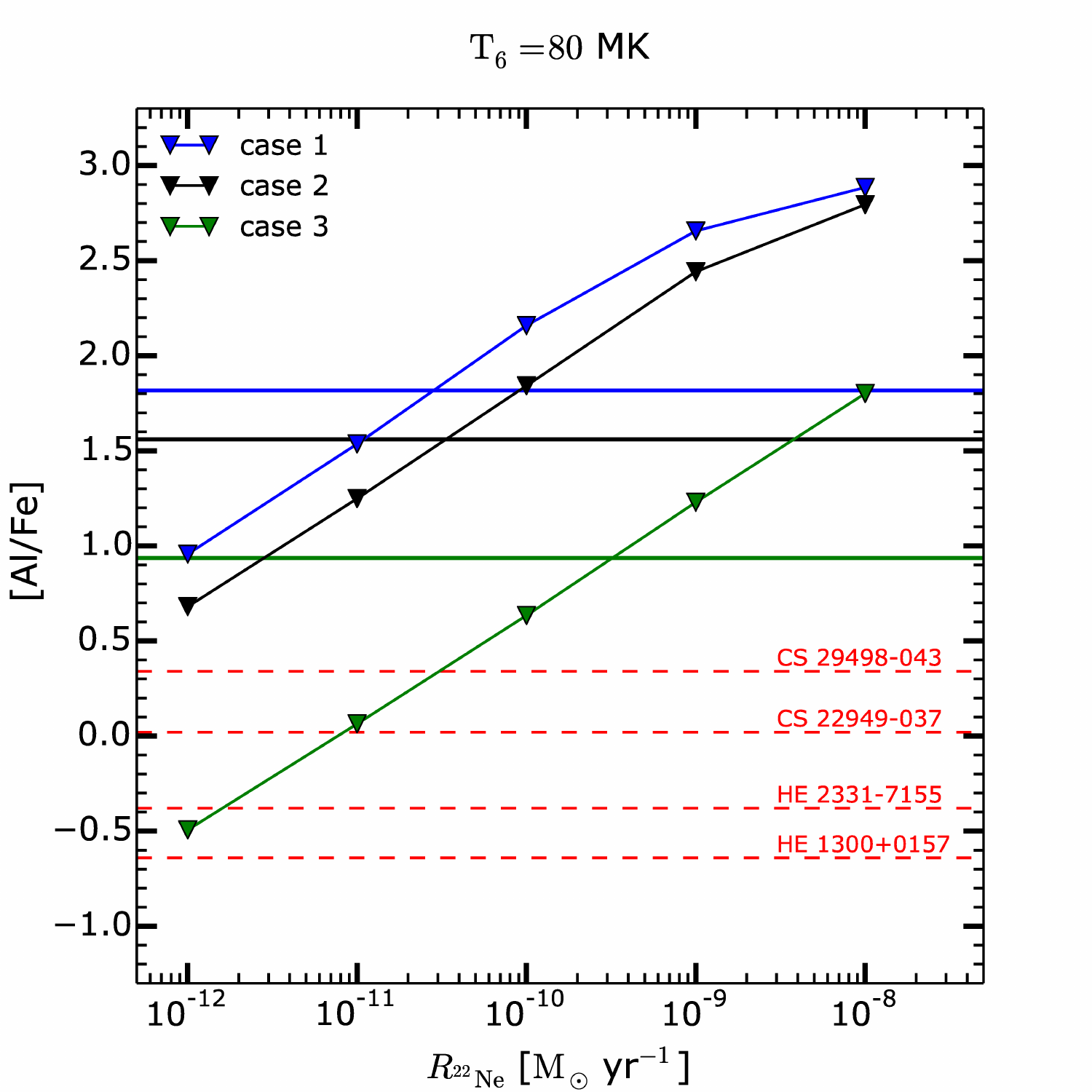}
   \end{minipage}
   \caption{\label{alfe} [Al/Fe] ratios in the H-box as a function of $R_{^{22}\rm Ne}$, the rate of injection of $^{22}$Ne ($^{12}$C and $^{16}$O are also injected). The temperature is $T_6=50$ MK (left) and $T_6=80$ MK (right), the density $\rho =$ 1 g cm$^{-3}$ and $^{12}$C, $^{16}$O and $^{22}$Ne are injected. The three lines with triangles show the final [Al/Fe] ratios in the H-box when considering three different sets of nuclear rates for the 3 principal reactions implying $^{27}$Al (see text for explanations). The horizontal blue, black and green lines corresponds to the initial [Al/Fe] in the H-box for the 3 considered cases. Observed [Al/Fe] are shown by the red dashed lines for the CEMP-no stars that have [Fe/H] = $ -3.8 \pm 0.3$.}
\end{figure*}

\section{The effects impacting the Al and Si abundances}
\label{sec:6}

The experiments presented in Sect. \ref{sec:4} show that the [Al/Fe] ratio in the H-box at hydrogen exhaustion lies always above the observed range of values. Even the initial [Al/Fe], taken from the hydrogen burning shell of a 60 $\rm M_{\odot}$ model at the beginning of the core helium burning phase is just above the observed range (see lines with squares on Fig. \ref{xfe}). We investigate now 3 effects impacting the aluminium abundance: the injection (species and rate), the temperature and the nuclear reaction rates.

\subsection{The impact of the injection on Al: species and rate}

First of all, one expects a lower final [Al/Fe] ratio if no $^{26}$Mg is injected. If it is injected, some $^{27}$Al is created through the reaction $^{26}$Mg($p,\gamma$)$^{27}$Al. This is illustrated in the third and fourth panels of Fig. \ref{xfe}. We see indeed that injecting $^{26}$Mg leads to a higher final [Al/Fe] than if no $^{26}$Mg is injected.

Fig. \ref{alfe} shows [Al/Fe] as a function of $R_{^{22}\rm Ne}$, the rate of injection of $^{22}$Ne at $T_6 = 50$ MK (left panel) and $T_6 = 80$ MK (right panel). The injected species are $^{12}$C, $^{16}$O and $^{22}$Ne.
The blue, black and green lines are associated to 3 sets of nuclear reactions rates that we tested and that will be discussed on Sect. \ref{rates}. Let us focus on the case 2 (black lines) that corresponds to the nuclear rates used until now. The horizontal black line shows the initial [Al/Fe] ratio in the H-box and the black line with triangles shows the final ones. 
We see that the final [Al/Fe] ratio at $T_6 = 80$ MK (right panel of Fig. \ref{alfe}) is lower when less $^{22}$Ne is injected. When the injection rate is low enough, the $^{27}$Al is globally more destroyed than created. We see that some aluminium is created at the end if $R_{^{22}\rm Ne} > $ 3 10$^{-11}$ $\rm M_{\odot}$ yr$^{-1}$ for the considered case. This aluminium comes mainly from the injected $^{22}$Ne thanks to successive proton captures.

Those two arguments suggest that a moderate amount of $^{22}$Ne coming from the helium burning core to the hydrogen burning shell together with no $^{26}$Mg would probably play in favor of a lower final [Al/Fe] ratio.

\subsection{The impact of the temperature on Al}

\label{impacttemp}

The final abundance of aluminium depends on the strength of the nuclear reactions rates that create and destroy it. The nuclear rates for the Ne-Na and Mg-Al cycles are generally 3-4 dex higher at $T_6 = 80$ MK than at $T_6 = 50$ MK so that the synthesis of aluminium is slower at $T_6 = 50$ MK. Fig. \ref{alfe} shows that indeed, the final [Al/Fe] ratios deviate only little from their initial values at $T_6 = 50$ MK (left panel) while the difference is much more significant at $T_6=80$ MK (right panel). Note however that the 'case 3' pattern at $T_6 = 50$ MK stands largely below its initial value (horizontal green line) but this is due to the $^{26}$Al that has decayed into $^{26}$Mg at the end of the simulation (this reduces [Al/Fe] by about 1 dex). Also the green triangle at the abscissa $10^{-12}$ $\rm M_{\odot}$ yr$^{-1}$ deviates from the others. The considered nuclear rates in that case disfavor the synthesis of $^{27}$Al (see discussion on Sect. \ref{rates}). This together with the low injection rate that imply a longer time before hydrogen exhaustion, allow a larger depletion of $^{27}$Al than at higher injection rates.

Also, we saw on Sect. \ref{sec:4} (see also Fig. \ref{rhoT}, left panel) that increasing the temperature leads to a lower final [Al/Fe] ratio (and a higher [Si/Fe]). This stays true as long as some $^{26}$Mg is injected: if $^{22}$Ne is injected but not $^{26}$Mg (as in Fig. \ref{alfe}), [Al/Fe] can be higher when increasing the temperature. It can be seen by comparing the left and right panels of Fig. \ref{alfe} at the abscissa $10^{-9}$ $\rm M_{\odot}$ yr$^{-1}$ for instance. In that case, no $^{26}$Mg is injected so that $^{27}$Al comes mainly from the successive proton captures on the injected $^{22}$Ne. The chain leading to $^{27}$Al is longer if starting from $^{22}$Ne than from $^{26}$Mg. In that chain, the rate of $^{24}$Mg($p,\gamma$)$^{25}$Al at $T_6 = 50$ MK is very low compared to the other reaction rates. This reaction tends to stop the chain at $T_6 = 50$ MK so that the final content in $^{27}$Al is generally close to the initial one, even with high injection rates of $^{22}$Ne.
Injecting some $^{26}$Mg is a way to avoid this bottleneck reaction and synthesize some aluminum, even at $T_6 = 50$ MK (see the fourth panel of Fig. \ref{xfe}). The rate of $^{24}$Mg($p,\gamma$)$^{25}$Al strongly increases from $T_6 = 50$ to 80 MK (by about 7 dex) so that the chain leading to $^{27}$Al (and $^{28}$Si) is no more blocked at $T_6 = 80$ MK.

A moderate temperature in the hydrogen shell ($\sim$ $50$ MK) is likely more compatible with a lower [Al/Fe] under various amounts of $^{22}$Ne coming from the helium core to the hydrogen shell. In the strong mixing case (high $R_{^{22}\rm Ne}$) and at high temperatures, this could lead to a very high [Al/Fe] ratio (Fig. \ref{alfe}, right panel).

\subsection{The impact of changing the nuclear rates on Al}
\label{rates}
A point that deserves to be investigated is the uncertainties of the nuclear rates. 
In a hydrogen burning region, the two reactions destroying $^{27}$Al are $^{27}$Al($p,\gamma$)$^{28}$Si and $^{27}$Al($p,\alpha$)$^{24}$Mg. The reaction which create $^{27}$Al is $^{26}$Mg($p,\gamma$)$^{27}$Al.
The 3 nuclear rates associated to these reactions are very uncertain at the considered temperatures. To illustrate this point, we compared 3 sets of nuclear rates for the 3 mentioned reactions involving $^{27}$Al. We used rates provided by the JINA REACLIB database \citep{cyburt10}.
\begin{itemize}
\item Case 1: the best scenario for the $^{27}$Al synthesis. We took the maximum rate (at $T_6=80$) for $^{26}$Mg($p,\gamma$)$^{27}$Al \citep{cyburt10}. The minimum rate was taken for both $^{27}$Al($p,\gamma$)$^{28}$Si and $^{27}$Al($p,\alpha$)$^{24}$Mg \citep{vanwormer94}.
\item Case 2: we used the rates taken for this work (see Sect. \ref{sec:3}).
\item Case 3: the best scenario for $^{27}$Al destruction. The minimum rate for $^{26}$Mg($p,\gamma$)$^{27}$Al is from \cite{angulo99} and the maximum rates for $^{27}$Al($p,\gamma$)$^{28}$Si and $^{27}$Al($p,\alpha$)$^{24}$Mg are from \cite{cyburt10}.
\end{itemize}
To be consistent, we computed 2 other complete stellar models from the ZAMS to core helium burning ignition with the 2 new sets of nuclear rates (case 1 and 3). Depending on the set of nuclear rates chosen, the initial abundances in the H-box are taken from the hydrogen burning shell of the corresponding stellar model. 
The 3 initial [Al/Fe] ratios taken in the H-box are shown by the blue (case 1), black (case 2) and green (case 3) horizontal lines. We see that the scatter is significant. This is because the Mg-Al cycle is already operating in the core of the complete stellar model during the Main-Sequence so that the aluminum abundance is affected if changing the nuclear rates. Depending on the set of nuclear rates, it finally leads to a different aluminum content in the hydrogen shell at core helium burning ignition, hence in the H-box.

The lines with triangles correspond to the final [Al/Fe] ratios in the H-box for the 3 cases. We verify that the rates considered in this work (case 2) lead to a final [Al/Fe] ratios in between the two extreme cases. Whatever the injection rate, at least 1.5 dex separates the blue from the green pattern, the green one giving lower [Al/Fe] since this is the case where $^{27}$Al is the most destroyed and the less synthesized. 
A word of caution: for the case 3, the abundance of $^{26}$Al is higher than the one of $^{27}$Al during the burning, so that [Al/Fe] is significantly affected when decaying the $^{26}$Al into $^{26}$Mg. For the cases 1 and 2, $^{27}$Al is more abundant so that decaying $^{26}$Al at the end reduces only a little [Al/Fe].

4 out of 5 CEMP-no stars with [Fe/H] $=-3.8 \pm 0.3$ have an observed [Al/Fe]. Those ratios are represented by the red dashed lines on Fig. \ref{alfe}.
The scatter of the observed [Al/Fe] ratio is well enough covered by the case 3 (green pattern) if relying on different values for $R_{\rm ^{22}Ne}$, the injection rate of $^{22}$Ne. If we select that set of rates, our model indicates that (i) the CEMP-no stars with [Al/Fe] $\gtrsim 0$ could be made of a material processed by hydrogen burning at relatively high temperatures (about 80 MK) because T $\sim 50$ MK would not lead to a high enough aluminium content (see left panel of Fig. \ref{alfe}) and (ii) only a high enough injection rate of $^{22}$Ne can account for Al-enhanced CEMP-no stars. The four stars considered here do not show such a high [Al/Fe] ratio (except CS 29498-043 however, but with a modest enhancement). This might tend to disfavor the progenitors where the mixing is really strong (very high initial velocity) and with a high temperature in the hydrogen shell ($\gtrsim 60$ $\rm M_{\odot}$ stars). The two left possibilities are either a high temperature ($\sim$ 80 MK) in the hydrogen shell but a weak mixing (a low $R_{\rm ^{22}Ne}$), or a moderate temperature in the hydrogen shell ($\sim$ 50 MK) with a weak to strong mixing.
A moderate temperature in the hydrogen shell is more likely achieved in $\sim$ 20 $\rm M_{\odot}$ progenitors ($30-60$ MK) rather than in $\sim$ 60 $\rm M_{\odot}$ ones ($30-80$ MK, see Fig. \ref{shell}).

\subsection{Sensitivity of the Si abundance}

The silicon abundance is also affected when changing the injection, temperature or nuclear reaction rates. In the explored range of parameters (temperature, injection rate and nuclear reaction rates, see Fig. \ref{alfe}) and for the injected species considered ($^{12}$C, $^{16}$O and $^{22}$Ne), the final [Si/Fe] ranges from 1 to 1.8.

At $T_6 = 50$ MK, the final [Si/Fe] ratio depends very weakly on the injection rate. This is because the Ne-Si chain is stopped by the $^{24}$Mg($p,\gamma$)$^{25}$Al reaction (see Sect. \ref{impacttemp}). In that case, the final [Si/Fe] ratio is almost equal to the initial value, which ranges from 1 (case 1, lowest rate for $^{27}$Al($p,\gamma$)$^{28}$Si) to 1.2 (case 3, highest rate for $^{27}$Al($p,\gamma$)$^{28}$Si).

At $T_6 = 80$ MK, the final abundance of silicon is always enhanced compared to the initial one. The $^{24}$Mg($p,\gamma$)$^{25}$Al reaction does not block the Ne-Si chain anymore so that the injected $^{22}$Ne synthesizes some $^{28}$Si. 
The final [Si/Fe] ratios range between 1.2 and 1.8.

Three CEMP-no stars with [Fe/H] $=-3.8 \pm 0.3$ have a measured [Si/Fe]. The values are 0.77, 0.82 and 0.87. Those values being closer to the results given by the model at $T_6 = 50$ MK (1 $<$ [Si/Fe] $<$ 1.2), it might indicate a moderate temperature in hydrogen shell of the progenitor ($T_6 \simeq 50$ MK), consistent with the discussion in Sect. \ref{rates}.

\section{The possible astronomical sources of classes 2 and 4 CEMP-no stars}
\label{sec:7}

Through the present work, we suggest that the high observed abundances of C together with that of N, O, Na and Mg at the surface of the CEMP-no stars are the signature of a mixing between the helium and hydrogen burning regions of the source star, during its nuclear life. What are the objects able to experience such a mixing process? In the framework of our results but also in a more global context, we speculate on the possible progenitors of classes 2 and 4 CEMP-no stars.

\subsection{AGB stars}

AGB stars are known contributors to s-process elements.
They are generally believed to be responsible for the enrichment, by mass transfer, in s-elements observed at the surface of the CEMP-s stars. In addition to the abundances, the models have to reproduce the period of the binary system for instance, which can give tight contraints but increase also the difficulty of finding models matching the observations \citep[see, e.g., ][]{abate15}. 

In AGB stars, there is a mixing between the two shells. It could also in principle be 
enhanced by rotation or at least rotation may change the chemical structure of the star at the beginning of the
AGB phase \citep[see the 7 $\rm M_{\odot}$ model in ][for instance]{meynet10}.

It seems however that there are at least 2 difficulties in this scenario to explain the CEMP-no stars. Firstly, the AGB stars are experiencing the s-process and by definition, the CEMP-no class is not (or weakly) enriched in s-elements. This feature would be difficult to explain relying on AGB stars. Secondly, it seems difficult to account for the CEMP-no stars having [Fe/H] $\lesssim -4$ with the AGB stars: such a low iron content is likely indicating that only the most massive objects, more massive than the AGB stars, have contributed.

\subsection{Faint supernovae, mixing and fallback}

\cite{tominaga14} discussed the scenario of faint supernovae from Pop. III stars with mixing and fallback. In such models, only the outer layers are ejected from the progenitor. It is indeed needed to explain the observed CNO abundance patterns, as well as the low $^{12}$C/$^{13}$C ratios. This is in line with the discussion about the mixing of the ejecta: we saw in Sect. \ref{sec:5} that no (or a few) material processed by helium burning coming from the source star should be mixed to the hydrogen rich envelope, when the nuclear life of the star is finished. In other words, mainly the hydrogen rich envelope of the progenitor would be needed to form the future CEMP-no star.

On the other hand, these models are generally non-rotating, leading to some difficulties in explaining the high nitrogen abundance observed in some CEMP-no stars without invoking an extra mixing process in the progenitor. Rotating models were however considered in \cite{takahashi14}, predicting higher N/Fe ratios in the ejecta, quite in line with the observed ones at the surface of two out of the three CEMP-no stars they considered (HE 0107-5240 and HE 1327-2326).

\subsection{Contribution of more than one source}
\cite{limongi03} proposed a two steps scenario: a normal $\sim$ 15 $\rm M_{\odot}$ supernova responsible for the iron-peak elements, followed by a fainter one experiencing strong fallback, coming from a $\sim$ 35 $\rm M_{\odot}$. The second progenitor, more massive, enriches the ISM in light elements: C, N, Na and Mg. These elements are produced thanks to a partial mixing between hydrogen and helium burning shells that can occur in $Z=0$ models, even in non rotating models.

What seems to not match with the CEMP-no stars of classes 2 and 4 is the predicted $^{12}$C/$^{13}$C ratio of 240 for this two steps model. The predicted [$^{12}$C/$^{13}$C] ratio is 0.4 and this is not compatible with the values of the CEMP-no stars considered here (see Fig. \ref{C12C13}, right panel). We note however that 3 out of 12 have only a lower limit for the $^{12}$C/$^{13}$C ratio (see Table \ref{table:1}) so that such a high predicted $^{12}$C/$^{13}$C could be consistent with these CEMP-no stars.

\subsection{The spinstars}

Several signatures of fast rotation at low metallicity were found over the past years. 
One strong signature is that the large nitrogen abundances as well as the low $^{12}$C/$^{13}$C ratios observed in normal Very Metal-Poor (VMP) stars are much better reproduced by low metallicity chemical evolution models if including fast rotators, also called spinstars \citep{chiappini06, chiappini08}. 
Because of the high rotation, the injection process we have investigated here operates in the spinstar and it could be a way to obtain a material enriched in C, N, O, Na and Mg together with a low $^{12}$C/$^{13}$C, that will ultimately form a CEMP-no star of class 2 or 4.

The spinstars and more generally the objects experiencing a mixing between hydrogen and helium burning regions appear as interesting candidates for being the classes 2 and 4 CEMP-no progenitors.

\section{Conclusions}
\label{concl}

We studied the possibility of forming CEMP-no stars with a material processed by hydrogen burning coming from the source star. We carried out nuclear experiments where the convective hydrogen burning shell of the source star was modeled by a hydrogen burning single zone (H-box). The mixing between the helium burning core and the hydrogen burning shell was mimicked by injecting the products of helium burning $^{12}$C, $^{16}$O, $^{22}$Ne and $^{26}$Mg in the H-box. $^{14}$N, $^{23}$Na, $^{24}$Mg and $^{27}$Al are synthesized when injecting those species in the hydrogen burning zone. The $^{12}$C/$^{13}$C ratio is constant under various densities, temperatures in the H-box, and also under various injection rates. The [Al/Fe] ratio in the hydrogen burning zone lies generally above the observations. 
Using different nuclear reaction rates found in the literature for the reactions involving $^{27}$Al leads to a better coverage of the observed [Al/Fe] scatter. The high observed [Al/Fe] ratios are reproduced at sufficiently high hydrogen burning temperature (80 MK) and if the injection rate of $^{22}$Ne is high enough. This might point toward a massive (high temperature) and fast rotating (high injection rate) progenitor.

Through this work, we suggest that the high observed abundances of light elements at the surface of the CEMP-no stars are the signature of a mixing between the helium and hydrogen burning regions of the progenitor, during its nuclear life.
It supports the CEMP-no star formation scenario of \cite{maeder15} for classes 2 and 4. This scenario states that those stars are made of a material processed by hydrogen burning only but where products of helium burning coming from the helium core of the source star diffused into the hydrogen burning shell thanks to the rotational mixing. This arrival of new elements boosts the nucleosynthesis in the hydrogen burning shell. 
Considerations on the $^{12}$C/$^{13}$C ratio confirmed that the CEMP-no stars of classes 2 and 4 are made of a material that was only processed by hydrogen burning in the source star. This corroborate the assumption stating that the CEMP-no stars formed mainly with the hydrogen rich envelope of the source star. The $^{12}$C/$^{13}$C ratio is highly sensitive to the burning region considered in the source star (hydrogen or helium burning region). It could be used to constrain the part which is expelled from the source star at the time of the supernova in order to reproduce the observed $^{12}$C/$^{13}$C ratio at the surface of the CEMP-no stars. 

The spinstars are interesting candidates for being the class 2 and 4 CEMP-no progenitors because of their rotation that induces exchanges of material between the hydrogen and helium burning regions. This is giving some support to the idea that the rotation played an important role in the early chemical evolution of galaxies.

\begin{acknowledgements}
The authors thank the anonymous referee who helped to improve this paper through very constructive remarks. This work was supported by the Swiss National Science Foundation (project number 200020-160119).
\end{acknowledgements}

\bibliographystyle{aa} 
\bibliography{biblio.bib}

   \section{Appendix}

Let us consider a box of initial mass $m=$ 1 $\rm M_{\odot}$ where $X_i$ denotes the mass fraction of the element $i$. We consider also a reservoir composed only of the element $e$, so that its mass fraction $X_e '$ in the reservoir is 1.
During a time $\Delta t$, we inject a mass
\begin{equation}\label{deltam}
\Delta m = R_e \Delta t
\end{equation}
from the reservoir into the box. $R_e$ is the injection rate of the element $e$ expressed in $\rm M_{\odot}$ yr$^{-1}$.
After the injection, the new mass fraction $X_e^{new}$ of the injected element in the box is
\begin{equation}
X_e^{new} = \frac{X_e m + X_e ' \Delta m}{m + \Delta m} = \frac{X_e m  + R_e \Delta t}{m + R_e \Delta t}.
\end{equation}
The new mass fraction of the other elements in the box can be expressed as
\begin{equation}
X_{i \neq e}^{new} = \frac{X_i m}{m + R_e \Delta t}.
\end{equation}
Note that the initially 1 $\rm M_{\odot}$ H-box is growing in mass due to the injection. Its final mass is generally similar to the initial one and does never exceed $\sim 1.3$ $\rm M_{\odot}$ for the presented results, which stays relatively close to $\rm 1 M_{\odot}$.

The point is now to estimate $R_e$ the injection rate. Let us consider the example of the carbon.
In stellar models, the primary $^{14}$N is synthesized  through the diffusion of $^{12}$C and $^{16}$O from the helium core to the hydrogen burning shell.
The secondary $^{14}$N is formed with the initial CNO elements in the star.
One can roughly quantify $M_{^{14}N}^{prim}$ the mass fraction of primary $^{14}$N formed during the core helium burning phase :

\begin{equation}
M_{^{14}N}^{prim} =  \left( \int_{0}^{M} X_{^{14}N}(M_r) \, \mathrm{d}M_r \right)_{Y_c = 0} - (X_{C,ini} + X_{N,ini} + X_{O,ini}) M
\end{equation}

where $Y_c$ is the central $^{4}$He mass fraction, $X_{^{14}N}(M_r)$ the mass fraction of $^{14}$N at coordinate $M_r$, $X_{C,ini}$, $X_{N,ini}$, $X_{O,ini}$ the mass fractions of the CNO elements at the ZAMS and $M$ the total mass of the star at the end of the core helium burning phase.
$M_{^{14}N}^{prim}$ is defined as the total amount of $^{14}$N in the star at core helium exhaustion minus the amount of $^{14}$N that can be formed with the initial CNO content (secondary $^{14}$N). 
We suppose that all the $^{12}$C and $^{16}$O diffusing from the helium core to the hydrogen shell are transformed into $^{14}$N. In that case, to get a mass $M_{^{14}N}^{prim}$ of primary nitrogen in the star at the end of core helium burning, one need an average injection rate of ($^{12}$C + $^{16}$O) in the hydrogen shell of
\begin{equation}
R_{^{12}C + ^{16}O} = \frac{M_{^{14}N}^{prim}}{\tau_{HeB}  }
\end{equation}
where $\tau_{HeB}$ is the duration of the core helium burning phase. 
For a 60 $\rm M_{\odot}$ model at $Z=10^{-5}$ and at 70 \% of the critical velocity at the ZAMS, we find $R_{^{12}C + ^{16}O} = 5$ $10^{-8}$ $\rm M_{\odot}$ yr$^{-1}$.

The amount of primary nitrogen synthesized (hence the value of $R_{^{12}C + ^{16}O}$) can change significantly depending on the rotation, the mass of the model or the prescription for the rotational mixing for instance. In the present work we consider $10^{-10}<R_{^{12}C}<10^{-6}$ $\rm M_{\odot}$ yr$^{-1}$ (the chosen values are $10^{-10}$, $10^{-8}$ and $10^{-6}$ $\rm M_{\odot}$ yr$^{-1}$). Also, we set $R_{^{12}C} = R_{^{16}O}$ and $R_{^{22}Ne} = R_{^{26}Mg} = R_{^{12}C} / 100$. The factor 100 between the two rates accounts at first order for the fact that $^{22}$Ne and $^{26}$Mg are $\sim$100 times less abundant than $^{12}$C and $^{16}$O in the helium burning core of a low metallicity massive stellar model, so that $\sim$100 times less $^{22}$Ne and $^{26}$Mg will enter by rotational mixing in the hydrogen burning shell.

\end{document}